\documentclass[twocolumn]{aastex63}
\usepackage{amsmath}
\usepackage{amstext}
\usepackage{apjfonts}
\usepackage{bm}
\usepackage{multirow}
\usepackage{hyperref}
\hypersetup{
            pdftitle={All-Sky Faint DA White Dwarf Spectrophotometric Standards for Astrophysical Observatories: The Complete Sample},
            pdfauthor={DAWD Folks}
            bookmarks=true,
            bookmarksnumbered=true,
            colorlinks=true,
            anchorcolor=blue,
            citecolor=blue,
            linkcolor=blue,
           }
\usepackage[figure,figure*]{hypcap}
\usepackage[graphicx]{realboxes}
\usepackage{xurl}
\graphicspath{{figures/}}

\newcommand*{\teff}{$T_{\rm eff}$}
\newcommand*{\logg}{$\log~g$}
\newcommand*{\av}{$A_V$}
\begin{document}

\shorttitle{All-Sky Faint DA WD Spectrophotometric Standards}
\author [0000-0002-5722-7199]{Tim Axelrod}
\affiliation{The University of Arizona, Steward Observatory, 933 North Cherry Avenue, Tucson, AZ 85721}
\author[0000-0002-6839-4881]{Abhijit Saha}
\affiliation{NSF's National Optical Infrared Astronomy Research Laboratory, 950 North Cherry Avenue, Tucson, AZ 85719}
\author[0000-0001-6685-0479]{Thomas Matheson}
\affiliation{NSF's National Optical Infrared Astronomy Research Laboratory, 950 North Cherry Avenue, Tucson, AZ 85719}
\author [0000-0002-7157-500X]{Edward W. Olszewski}
\affiliation{The University of Arizona, Steward Observatory, 933 North Cherry Avenue, Tucson, AZ 85721}
\author[0000-0001-9806-0551]{Ralph C. Bohlin}
\affiliation{Space Telescope Science Institute, 3700 San Martin Drive, Baltimore, MD 21218}

\author[0000-0002-0882-7702]{Annalisa Calamida}
\affiliation{Space Telescope Science Institute, 3700 San Martin Drive, Baltimore, MD 21218}
\author{Jenna Claver}
\affiliation{NSF's National Optical Infrared Astronomy Research Laboratory, 950 North Cherry Avenue, Tucson, AZ 85719}
\author[0000-0003-2823-360X]{Susana Deustua}
\affiliation{Sensor Science Division, National Institute of Standards and Technology, Gaithersburg, MD 20899-8441}
\affiliation{Space Telescope Science Institute, 3700 San Martin Drive, Baltimore, MD 21218}
\author [0000-0003-3082-0774]{Jay B. Holberg}
\affiliation{The University of Arizona, Lunar and Planetary Laboratory, 1629 East University Boulevard, Tucson, AZ 85721}
\author [0000-0001-8816-236X]{Ivan Hubeny}
\affiliation{The University of Arizona, Steward Observatory, 933 North Cherry Avenue, Tucson, AZ 85721}
\author[0000-0001-6529-8416]{John W. Mackenty}
\affiliation{Space Telescope Science Institute, 3700 San Martin Drive, Baltimore, MD 21218}
\author [0000-0001-7179-7406]{Konstantin Malanchev}
\affiliation{University of Illinois at Urbana-Champaign, 1002 W. Green St., Urbana IL 61801}
\author[0000-0001-6022-0484]{Gautham Narayan}
\affiliation{University of Illinois at Urbana-Champaign, 1002 W. Green St., Urbana IL 61801}
\author [0000-0002-4596-1337]{Sean Points}
\affiliation{Cerro Tololo Inter-American Observatory, Casilla 603, La Serena, Chile}
\author[0000-0002-4410-5387]{Armin Rest}
\affiliation{Space Telescope Science Institute, 3700 San Martin Drive, Baltimore, MD 21218}
\affiliation{Department of Physics and Astronomy, Johns Hopkins University, Baltimore, MD 21218, USA}
\author [0000-0003-2954-7643]{Elena Sabbi}
\affiliation{Space Telescope Science Institute, 3700 San Martin Drive, Baltimore, MD 21218}
\author[0000-0003-0347-1724]{Christopher W. Stubbs}
\affiliation{Harvard University, Department of Physics, 17 Oxford Street, Cambridge, MA 02138}
\affiliation{Harvard-Smithsonian Center for Astrophysics, 60 Garden Street, Cambridge, MA 02138}

\defcitealias{Bergeron1992}{B92}
\defcitealias{Fitzpatrick99}{F99}
\defcitealias{Bohlin2014}{B14}
\defcitealias{Narayan16}{N16}
\defcitealias{Narayan19}{N19}
\defcitealias{Hermes17} {H17}
\defcitealias{Toonen17}{T17}
\defcitealias{Calamida19}{C19}
\defcitealias{Calamida22}{C22}
\defcitealias{Supercal}{S15}

\title{All-Sky Faint DA White Dwarf Spectrophotometric Standards for Astrophysical Observatories: The Complete Sample}
\begin{abstract}
Hot DA white dwarfs have fully radiative pure hydrogen atmospheres that are the least complicated to model.
Pulsationally stable, they are fully characterized by their effective temperature \teff, and surface gravity \logg,
which can be deduced from their optical spectra and used in model atmospheres to predict their spectral energy
distribution (SED). Based on this, three bright DAWDs have defined the spectrophotometric flux scale of the
CALSPEC system of HST. In this paper we add 32 new fainter (16.5 < V < 19.5) DAWDs spread over the
whole sky and within the dynamic range of large telescopes. Using ground based spectra and panchromatic
photometry with HST/WFC3, a new hierarchical analysis process demonstrates consistency between model and
observed fluxes above the terrestrial atmosphere to < 0.004 mag rms from 2700 Å to 7750 Å and to 0.008
mag rms at 1.6µm for the total set of 35 DAWDs. These DAWDs are thus established as spectrophotometric
standards with unprecedented accuracy from the near ultraviolet to the near-infrared, suitable for both ground
and space based observatories. They are embedded in existing surveys like SDSS, PanSTARRS and GAIA,
and will be naturally included in the LSST survey by Rubin Observatory. With additional data and analysis to
extend the validity of their SEDs further into the IR, these spectrophotometric standard stars could be used for JWST, as well as for the Roman and Euclid observatories.
\end{abstract}    
 

\section{Introduction}
\label{sec:Intro}

Most currently available spectrophotometric (and photometric) standards in the sky limit us to color accuracies of 1 to 2\%. This accuracy is a limitation to the uncertainty budgets of key scientific investigations such as the determination of photo-redshifts. This in turn limits the uncertainties in determining the dark-energy equation of state, e.g.~\citet{Betoule13}. The motivation for developing an all-sky network of DA white dwarfs (DAWDs) as more accurate spectrophotometric standards was described in considerable detail by \citet[hereafter N16]{Narayan16}, and need not be repeated here. Salient features of those arguments are briefly re-cast in \S~\ref{sec:ratmetevo} below.

This paper presents an all-sky set of 32 new spectrophotometric standard stars on an absolute scale. They are faint enough to be within the dynamic range of large telescopes (apertures 4m and higher), with two or more of them accessible from any site on the globe at any instant at airmass lower than 2. 
This study has utilized observational data from the {\it Hubble Space Telescope (HST)} through three proposals: GO-12967, GO-13711, and GO-15113 (PI: A. Saha), and spectroscopic observations from the ground utilizing data from Gemini Observatory, the MMT Observatory, and the SOAR telescope. 

In prior publications we have presented results for a sample of 19 stars in the equatorial and northern regions of the sky with spectrophotometric accuracy in colors to sub-percent accuracy from the near ultraviolet (UV) through near infrared (IR). In this paper we add an additional 13
DAWDs in the Southern sky to extend the 19 Northern and Equatorial faint DAWD standards presented in \citet[hereafter N19]{Narayan19}, thereby yielding an all-sky network of 32 faint spectrophotometric standards.  These 13 Southern standards (observed in HST Cycle 25) are analyzed using the same technique developed in \citetalias{Narayan19}
for observations from Cycles 20 and 22. Results from this analysis are then input to a new simultaneous analysis of the entire all-sky network of the 32 faint stars and 
three CALSPEC\footnote{\url{http://www.stsci.edu/hst/instrumentation/reference-data-for-calibration-and-tools/astronomical catalogs/CALSPEC}} standards \citep{Bohlin2014a}.  For brevity, the exposition of the \citetalias{Narayan19} analysis method is incorporated only by reference. Here, the emphasis is on the new Southern standards and the new simultaneous analysis method.

\S\ref{sec:ratmetevo} recapitulates the motivation for establishing faint white dwarfs as spectrophotometric standards, as well as the concepts that underpin our approach to doing so. We outline how our analysis processes have evolved through our past publications and led to our complete ``final'' sample. Our calibration stands maximally independent of all other spectrophotometric systems and standards and is dependent only on how well the atmospheres of pure hydrogen white dwarfs can be modeled.
\S\ref{sec:targsel} briefly describes the selection process for the new Southern DAWDs, referring extensively to 
\citetalias{Narayan16} and \citetalias{Narayan19}, as well as \citet[hereafter C19]{Calamida19} and \citet[hereafter C22]{Calamida22}.   \S\ref{sec:data} describes the observations and their reductions, all of which are closely similar to \citetalias{Narayan19}.  \S\ref{sec:danaly} presents the \citetalias{Narayan19} data reduction scheme as applied to the Southern candidates, while \S\ref{sec:analys-a} and \S\ref{sec:analys-b} present the simultaneous analysis of all Northern, equatorial, and Southern standards. \S\ref{sec:results} compares the results to CALSPEC and gives calculated magnitudes for DES, DECaLS, PaNSTARRS DR1, SDSS (DR 7), and Gaia (DR3). Finally, \S\ref{sec:concl} presents the conclusions. 

\section{Rationale, Methodology, and Evolution}
\label{sec:ratmetevo}
If a DA white dwarf is hotter than $\sim$15000K, the atmosphere is completely radiative, making it the least complicated type of star to model. Such stars are characterizable by two parameters, \teff\ and \logg\ \citep{Holbergetal1985}. The shapes and widths of the Balmer features, and the Balmer jump when available, determine these two parameters. An atmospheric model with these parameters then predicts the emergent spectral energy distribution (SED) for the DA white dwarf. Measurements of the incident SED made above the terrestrial atmosphere (by an instrument whose response stability can be independently monitored) can be used to verify if the model predicted SEDs agree with the measurements.

Investigations by \citet{Bohlin2014a, Bohlin2020} and \citet{Bohlin2022}, using HST and its instruments, provided empirical verification of this concept. Three DA white dwarfs with $V \approx 12$ mag were established as SED standards covering  UV, visible and near-IR wavelengths. These stars define the HST CALSPEC system. The relative flux vs. wavelength of these stars is based on the physical properties of the stars and rests on our understanding of the physics of their atmospheres, independently of the absolute flux level, which depends only on the absolute monochromatic flux of Vega at 5557.5~\AA\ (vacuum) and the IR flux of Sirius \citep{Bohlin2020}. 

To create standards in the high signal-to-noise ratio (S/N) range, but not saturated in observations from  telescopes of up to 30m aperture, the standards must be fainter than $V\ge$16.5~mag, with a median brightness even fainter, say, $V=18$~mag. This implies that they must be $\approx 10$ times more distant than Bohlin's triad that define the CALSPEC system. Unfortunately, this also puts them at large enough distances that interstellar extinction can no longer be ignored. The comparison of model predicted SEDs with observations can account for this by allowing and solving for the extinction \av\ of each individual star when comparing observations with prediction (assuming that the total to selective extinction characterized by $R_{V}$ is the same for all the stars). Thus there are now three parameters that quantitatively characterize the SED received above the terrestrial atmosphere: \teff, \logg, and \av, plus an overall achromatic normalization that scales the absolute brightness at all wavelengths.

In \citetalias{Narayan16} these precepts were directly applied to four stars: the details are available there and are not repeated here. There were several areas in which improvement was desirable:
\begin{itemize}
    \item 
    The parameters \teff\ and \logg\ are determined in a separate step from the one for \av\ so that errors from the first propagate into the determination of \av\ which is undesirable.
    \item
    Given that the Balmer lines can be very broad and vary from object to object depending on its temperature and pressure, locating the continuum can be subjective, inducing errors in determining the atmospheric model parameters. 
\end{itemize}

\citetalias{Narayan19} introduced a hierarchical analysis that addresses the first issue, while utilizing a Gaussian process model for the flux calibration errors to mitigate the second. Figure 16 of \citetalias{Narayan19} shows 1-$\sigma$ residuals of 0.003 to 0.005 mag, except for the F160W band, which had mean of 0.009 mag and 1-$\sigma$ of 0.013 mag. The reader is referred to the extensive discussion in \citetalias{Narayan19} for details. 

In \citetalias{Narayan19}, each of the stars is treated individually, with zero-points in each passband determined by the CALSPEC calibration. It is instead possible to reduce the CALSPEC observations and observations of the 13 Southern and 19 Northern/Equatorial simultaneously, the equivalent of Bohlin's experiment, which is to see how self-consistent the predicted SEDs are for all the final selected candidates (including also the three CALSPEC standards), without reference to any pre-existing calibration. This allows the relative band to band zero-point differences to be determined by the entire ensemble of stars, which potentially sets accuracies in colors more robustly than the CALSPEC calibration. The only unknown left is the monochromatic zero-point scalar that adjusts the flux level equally in all bands to match the canonical brightness of Vega at a reference wavelength of 5557.5~\AA\ (vac) according to \citet{Megessier95}, as adjusted to $3.44\times10^{-9}$~erg cm$^{-2}$ s$^{-1}$ \AA$^{-1}$ by \citet{Bohlin2014a}. Our next paper will
adopt the revised value of $3.47\times10^{-9}$~erg cm$^{-2}$ s$^{-1}$ \AA$^{-1}$ \citep{Bohlin2020}.
Details are in \S\ref{sec:analys-a} and \S\ref{sec:analys-b}.


\section{Target selection}
\label{sec:targsel}

\citetalias{Narayan19} and \citetalias{Calamida19} picked known probable DAWDs from SDSS observations \citep{Kleinmannetal2004, Eisensteinetal2006} and from the Villanova \citep{vill1999} catalog\footnote{\url{http://www.astronomy.villanova.edu/WDCatalog/}}, which is now superseded by
the Montreal \citep{montreal2017} WD database\footnote{\url{ https://www.montrealwhitedwarfdatabase.org/}}. All of these northern and equatorial candidates had published low-resolution spectra whose quality was sufficient to decide if they were DA, if they were hot enough to be fully radiative, and if there were no obvious issues such as magnetic line splitting or trace atmospheric elements. 
We observed all these Northern/Equatorial stars at Gemini (GMOS, 1\farcs5 or 1\farcs0 slit, 0.92 \AA/mm, 3500-6360\AA\  coverage) and/or at the MMT (Blue channel, 300 line grating, 1\farcs0 or 1\farcs25 slits, 1.95 \AA/mm, 3400-8400\AA\  coverage).
These spectra, their reduction and analysis are presented in \citetalias{Calamida19}.

When searching for southern stars, there was no obvious equivalent list of faint WDs, so candidates were selected from the Supercosmos and VST surveys 
\citep{Gentile-Fusilloetal2017, Raddietal2016, Raddietal2017} by using photometry and proper motion selection criteria (absolute magnitude brighter than 9.0). For more details please see \citetalias{Calamida22}.

\section {Data collection, SOAR spectroscopy, HST photometry, and reduction}
\label{sec:data}

Approximately 50 southern WD candidates were observed using the Goodman spectrograph at SOAR (1\farcs07 slit, 1.99 A/pixel, 3850-7100A coverage) over several runs in 2016 and 2017. A list of candidates observed and a log of observations are in \citetalias{Calamida22}. To be conservative, the final list includes 15 DAWDs (two rejected in the next section) with \teff\ > 20000K observed with a total of three Cycle 25 HST orbits per star.  The three primary CALSPEC spectrophotometric standards, GD 71, GD 153, and G191-B2B, have three visits each in HST Cycle 25 to mitigate possible WFC3 sensitivity changes.

Spectral reductions were completed as in \citetalias{Narayan19}. Using the machinery in \citetalias{Narayan19}, we derived \teff, \logg, and \av, which, as will be discussed in \S\ref{sec:analys-a}, become input priors for the new analysis (along with their error distributions).

Final photometric reductions, as described in \citetalias{Calamida19}, were performed with Saha's ILAPH.  This custom aperture photometry code offers interactive “growth curve” analysis for optimal sky subtraction. In addition to extracting the best possible count-rates from the images, it is critical to ensure that they are all on a fully self-consistent system of instrumental magnitudes.  Since our {\it HST/WFC3} data were acquired at different times over several years, and with different instrument configurations, special attention was paid to adjusting for possible systematic shifts. These are described in considerable detail in \citetalias{Calamida19} and \citetalias{Calamida22}. The end result, presented in 
Table  \ref{tab:HST_Photometry}, puts the photometry on the {\it existing} AB magnitude system of CALSPEC. The fact that these are AB magnitudes is irrelevant {\it per se} for subsequent analysis, except that it was a convenient way to ensure that they are based on a self-consistent instrumental system.  The eventual result is the derivation of new AB magnitudes that are not dependent on this particular starting point for the observed magnitudes.

\begin{deluxetable*}{llccccccc}
\tabletypesize{\scriptsize}
\tablecaption{\emph{Gaia} DR3 astrometry and photometry for the candidate spectrophotometric standard DA white dwarfs\label{tab:Targets}}
\tablehead{
\colhead{Star}&
\colhead{Orig. name}&
\colhead{RA\tablenotemark{a}}&
\colhead{DEC\tablenotemark{a}}&
\colhead{$PM_{RA}$}&
\colhead{$PM_{DEC}$}&
\colhead{$G$}&
\colhead{$Rp$}&
\colhead{$Bp$}\\
\colhead{}&
\colhead{}&
\colhead{(hh:mm:ss.s)}&
\colhead{(dd:mm:ss.s)}&
\colhead{(mas/yr)}&
\colhead{(mas/yr)}&
\colhead{mag}&
\colhead{mag}&
\colhead{mag}
}
\startdata
\multicolumn{9}{c}{Northern and equatorial DAWDs} \\
\hline                    
WDFS0103-00 & SDSSJ010322.19-002047.7	   & 01:03:22.201 & $-$00:20:47.800   &   \phantom{$-$1}6.196$\pm$0.382  &  \phantom{1}$-$6.550$\pm$0.355 & 19.30 & 19.67 & 19.16 \\
WDFS0228-08 & SDSSJ022817.16-082716.4	   & 02:28:17.183 & $-$08:27:16.301 &  \phantom{$-$}10.916$\pm$0.783  &   \phantom{$-$1}3.151$\pm$0.539 & 19.97 & 20.07 & 19.82 \\
WDFS0248+33 & SDSSJ024854.96+334548.3	   & 02:48:54.965 &  \phantom{$-$}33:45:48.244 &   \phantom{$-$1}4.093$\pm$0.253  &  \phantom{1}$-$4.759$\pm$0.205 & 18.52 & 18.74 & 18.42 \\
  \ldots    & SDSSJ041053.632-063027.580\tablenotemark{*} & \phantom{$-$}04:10:53.641 & $-$06:30:27.677 & \phantom{$-$1}8.577$\pm$0.279  & \phantom{$-$1}9.719$\pm$0.185 & 18.99 & 19.22 & 19.02  \\
  \ldots    & WD0554-165\tablenotemark{*}                 & \phantom{$-$}05:57:01.292 & $-$16:35:12.159 & \phantom{1}$-$6.747$\pm$0.099 & \phantom{$-$1}4.272$\pm$0.101 & 17.94 & 18.40 & 17.83  \\
WDFS0727+32 & SDSSJ072752.76+321416.1	   & 07:27:52.752 &  \phantom{$-$}32:14:16.046 & $-$13.151$\pm$0.168  &  \phantom{1}$-$6.923$\pm$0.128 & 18.19 & 18.45 & 18.04  \\
WDFS0815+07 & SDSSJ081508.78+073145.7	   & 08:15:08.782 &  \phantom{$-$}07:31:45.775 &   \phantom{$-$1}5.519$\pm$0.811  &  \phantom{1}$-$0.190$\pm$0.733 & 19.93 & 20.25 & 19.79  \\
WDFS1024-00 & SDSSJ102430.93-003207.0	   & 10:24:30.912 & $-$00:32:07.16  & $-$21.301$\pm$0.388  &  \phantom{1}$-$5.670$\pm$0.590 & 19.08 & 19.23 & 19.00  \\
WDFS1110-17 & SDSSJ111059.42-170954.2	   & 11:10:59.436 & $-$17:09:54.308 &   \phantom{$-$1}5.454$\pm$0.162  &  \phantom{1}$-$8.015$\pm$0.136 & 18.05 & 18.37 & 17.91  \\
WDFS1111+39 & SDSSJ111127.30+395628.0	   & 11:11:27.313 &  \phantom{$-$}39:56:28.105 &   \phantom{$-$1}2.734$\pm$0.231  &   \phantom{$-$1}2.933$\pm$0.255 & 18.64 & 19.07 & 18.48  \\
WDFS1206+02 & SDSSJ120650.504+020143.810   & 12:06:50.41  &  \phantom{$-$}02:01:42.138 &  \phantom{1}$-$5.061$\pm$0.300  & $-$23.367$\pm$0.149 & 18.85 & 19.07 & 18.75  \\
WDFS1214+45 & SDSSJ121405.11+453818.5	   & 12:14:05.111 &  \phantom{$-$}45:38:18.626 &   \phantom{$-$1}0.278$\pm$0.088  &  \phantom{$-$}13.925$\pm$0.104 & 17.98 & 18.23 & 17.84  \\
WDFS1302+10 & SDSSJ130234.43+101238.9	   & 13:02:34.422 &  \phantom{$-$}10:12:38.717 & $-$12.856$\pm$0.132  & $-$16.837$\pm$0.122 & 17.24 & 17.54 & 17.10  \\
WDFS1314-03 & SDSSJ131445.050-031415.588   & 13:14:45.046 & $-$03:14:15.685 &  \phantom{1}$-$3.930$\pm$0.404  &  \phantom{1}$-$5.659$\pm$0.265 & 19.31 & 19.74 & 19.25  \\
WDFS1514+00 & SDSSJ151421.27+004752.8	   & 15:14:21.277 &  \phantom{$-$}00:47:52.380  &   \phantom{$-$1}4.350$\pm$0.059  & $-$26.855$\pm$0.053 & 15.88 & 16.11 & 15.77  \\
WDFS1557+55 & SDSSJ155745.40+554609.7	   & 15:57:45.38  &  \phantom{$-$}55:46:09.361 & $-$11.677$\pm$0.112  & $-$21.478$\pm$0.126 & 17.69 & 18.04 & 17.53  \\
WDFS1638+00 & SDSSJ163800.360+004717.822   & 16:38:00.352 &  \phantom{$-$}00:47:17.739 &  \phantom{1}$-$9.171$\pm$0.320  &  \phantom{1}$-$2.737$\pm$0.239 & 19.02 & 19.36 & 18.91  \\
  \ldots    & SDSSJ172135.97+294016.0\tablenotemark{*} & 17:21:35.951 & \phantom{$-$}29:40:16.178 & $-$20.919$\pm$0.230 & \phantom{$-$}10.536$\pm$0.260 & 19.60 & 19.50    & 19.69  \\
WDFS1814+78 & SDSSJ181424.075+785403.048   & 18:14:24.078 &  \phantom{$-$}78:54:03.084 & $-$10.738$\pm$0.060 & \phantom{$-$}11.535$\pm$0.057  & 16.74  & 17.03 & 16.61  \\
  \ldots    & SDSSJ203722.169-051302.964\tablenotemark{*} & 20:37:22.173 & $-$05:13:03.023 & \phantom{$-$1}3.118$\pm$0.267 & \phantom{1}$-$2.000$\pm$0.206 & 19.11 & 19.40 & 19.04 \\
WDFS2101-05 & SDSSJ210150.65-054550.9	   & 21:01:50.667 & $-$05:45:51.159 &   \phantom{$-$1}9.984$\pm$0.218  & $-$11.694$\pm$0.210 & 18.83 & 19.10 & 18.74  \\
WDFS2329+00 & SDSSJ232941.330+001107.755   & 23:29:41.321 &  \phantom{$-$}00:11:07.565 &  \phantom{1}$-$7.982$\pm$0.189  & $-$14.919$\pm$0.162 & 18.29 & 18.42 & 18.24  \\
WDFS2351+37 & SDSSJ235144.29+375542.6	   & 23:51:44.274 &  \phantom{$-$}37:55:42.569 & $-$16.412$\pm$0.145  &  \phantom{1}$-$9.941$\pm$0.107 & 18.23 & 18.50 & 18.12  \\
\hline                    
\multicolumn{9}{c}{Southern DAWDs} \\
\hline                    
WDFS0122-30 & A020.503022 & 01:22:00.725 & $-$30:52:03.95    &  \phantom{$-$}20.621$\pm$0.14  & $-$12.303$\pm$0.135 & 18.66 & 19.01 & 18.53  \\
WDFS0238-36 & SSSJ023824  & 02:38:24.969 & $-$36:02:23.222   &  \phantom{$-$}57.993$\pm$0.078 &  \phantom{$-$}13.747$\pm$0.119 & 18.24 & 18.39 & 18.19  \\
\ldots      & WD0418-534\tablenotemark{*} & 04:19:24.68 & $-$53:19:16.659 & $-$17.587$\pm$0.048 & \phantom{$-$}27.166$\pm$0.063 & 16.42 & 16.69 & 16.30  \\
WDFS0458-56 & SSSJ045822  & 04:58:23.133 & $-$56:37:33.434   & 143.596$\pm$0.118 &  \phantom{$-$}66.486$\pm$0.130  & 17.96 & 18.25 & 17.85  \\
WDFS0541-19 & SSSJ054114  & 05:41:14.759 & $-$19:30:38.896   &  \phantom{$-$}19.248$\pm$0.126 & $-$26.954$\pm$0.142 & 18.43 & 18.61 & 18.35  \\
WDFS0639-57 & SSSJ063941  & 06:39:41.468 & $-$57:12:31.164   &  \phantom{$-$}17.513$\pm$0.126 &  \phantom{$-$}43.576$\pm$0.151 & 18.37 & 18.70 & 18.27  \\
\ldots      & WD0757-606\tablenotemark{*} & 07:57:50.637& $-$60:49:54.634 &  \phantom{1}$-$4.590$\pm$0.287 & \phantom{$-$}11.067$\pm$0.223 & 18.95 & 19.15 & 18.89  \\
WDFS0956-38 & SSSJ095657  & 09:56:57.009 & $-$38:41:30.269   &  \phantom{1}$-$8.269$\pm$0.084 & $-$46.075$\pm$0.092 & 18.00 & 18.16 & 17.94\ \\
WDFS1055-36 & SSSJ105525  & 10:55:25.356 & $-$36:12:14.731   & $-$21.353$\pm$0.124 &  \phantom{$-$}46.134$\pm$0.119 & 18.20 & 18.45 & 18.12  \\
WDFS1206-27 & WD1203-272  & 12:06:20.354 & $-$27:29:40.639   &   \phantom{$-$1}3.019$\pm$0.074 &   \phantom{$-$1}2.796$\pm$0.081 & 16.67 & 16.93 & 16.54  \\
WDFS1434-28 & SSSJ143459  & 14:34:59.528 & $-$28:19:03.295   & $-$48.559$\pm$0.206 &  \phantom{$-$}18.600$\pm$0.195 & 18.10 & 18.35 & 18.07  \\
WDFS1535-77 & WD1529-772  & 15:35:45.179 & $-$77:24:44.832   & $-$26.881$\pm$0.055 & $-$43.749$\pm$0.058 & 16.76 & 17.09 & 16.60  \\
WDFS1837-70 & SSSJ183717  & 18:37:17.906 & $-$70:02:52.513   &  \phantom{$-$}10.378$\pm$0.072 & $-$75.989$\pm$0.106 & 17.91 & 18.08 & 17.85  \\
WDFS1930-52 & SSSJ193018  & 19:30:18.995 & $-$52:03:46.55    &  \phantom{$-$}21.546$\pm$0.123 & $-$33.286$\pm$0.102 & 17.67 & 17.94 & 17.55  \\
WDFS2317-29 & WD2314-293  & 23:17:20.294 & $-$29:03:21.647   &   \phantom{$-$1}3.991$\pm$0.146 &  \phantom{$-$}25.051$\pm$0.196 & 18.53 & 18.81 & 18.44  \\
\enddata
\tablenotetext{a}{Coordinates are from \emph{Gaia} DR3 at epoch J2016.0.}
\tablenotetext{*}{This star was excluded from the final network of spectrophotometric standard DAWDs. See text of \citetalias{Calamida22} for more details. }
\end{deluxetable*}

\subsection{Las Cumbres Time Series Photometry}

The main purposes of the work presented in \citetalias{Calamida22} were to show spectra of all the stars observed with SOAR and to test these stars for variability using the Las Cumbres Observatory (LCO) network of telescopes. While hot DAWDs are not expected to be intrinsically variable, they could be variable because of binary companions,``seeing variables'' due to close faint red stars, or dust cloud remnants around the WD. \citetalias{Narayan19} and \citetalias{Calamida19} discuss the rejection of 4 candidate stars for spectroscopic and photometric reasons,
while  \citetalias{Calamida22}  rejects a total of six stars in the all-sky set
leaving 32 faint stars, plus the three brighter CALSPEC standards, to form our network.
The details of the resulting set of target stars are in Table \ref{tab:Targets}, duplicated from \citetalias{Calamida22}, which contains further details of our target selection procedure.

\begin{deluxetable*}{ccccccccccccc}
\scriptsize
\tablecaption{HST observed photometry. Units are AB mag.\tablenotemark{a}\label{tab:HST_Photometry}}
\tablehead{
\colhead{Object} & \colhead{F275W} & \colhead{cF275W} & \colhead{F336W} & \colhead{cF336W} & \colhead{F475W} & \colhead{cF475W} & \colhead{F625W} &  \colhead{cF625W} & \colhead{F775W}  & \colhead{cF775W}  & \colhead{F160W} & \colhead{cF160W}}
\startdata
G191B2B &  10.490(1)\phantom{1} & 10.494 &  10.890(1)\phantom{1} & 10.892 &  11.499(1) & 11.498 &  12.031(1) & 12.030 &  12.451(1)\phantom{1} & 12.451 &  13.885(2)\phantom{1}   & 13.883 \\
GD153 &  12.202(2)\phantom{1} & 12.205 &  12.568(1)\phantom{1} & 12.570 &  13.100(2) & 13.099 &  13.598(1) & 13.597 &  14.002(1)\phantom{1} & 14.002 &  15.414(2)\phantom{1}   & 15.409 \\
GD71 &  11.989(1)\phantom{1} & 11.992 &  12.336(1)\phantom{1} & 12.338 &  12.799(1) & 12.798 &  13.279(1) & 13.278 &  13.672(1)\phantom{1} & 13.672 &  15.068(2)\phantom{1}   & 15.063 \\
WDFS0103-00 &  18.195(4)\phantom{1} & 18.199 &  18.527(5)\phantom{1} & 18.529 &  19.083(5) & 19.082 &  19.569(5) & 19.568 &  19.965(6)\phantom{1} & 19.965 &  21.355(12)   & 21.340 \\
WDFS0122-30 &  17.671(3)\phantom{1} & 17.674 &  17.994(3)\phantom{1} & 17.997 &  18.460(3) & 18.459 &  18.922(3) & 18.921 &  19.320(3)\phantom{1} & 19.320 &  20.705(7)\phantom{1}   & 20.691 \\
WDFS0228-08 &  19.518(8)\phantom{1} & 19.522 &  19.715(10) & 19.718 &  19.815(7) & 19.814 &  20.169(7) & 20.168 &  20.501(6)\phantom{1} & 20.501 &  21.737(17)   & 21.721 \\
WDFS0238-36 &  17.790(3)\phantom{1} & 17.794 &  17.972(2)\phantom{1} & 17.974 &  18.095(2) & 18.094 &  18.439(3) & 18.438 &  18.757(3)\phantom{1} & 18.757 &  19.992(5)\phantom{1}   & 19.979 \\
WDFS0248+33 &  17.829(4)\phantom{1} & 17.832 &  18.040(6)\phantom{1} & 18.042 &  18.370(3) & 18.369 &  18.746(3) & 18.745 &  19.077(2)\phantom{1} & 19.077 &  20.340(6)\phantom{1}   & 20.326 \\
WDFS0458-56 &  17.023(2)\phantom{1} & 17.027 &  17.351(3)\phantom{1} & 17.353 &  17.754(3) & 17.754 &  18.217(2) & 18.216 &  18.601(2)\phantom{1} & 18.601 &  19.999(5)\phantom{1}   & 19.987 \\
WDFS0541-19 &  18.021(3)\phantom{1} & 18.024 &  18.215(3)\phantom{1} & 18.218 &  18.276(2) & 18.275 &  18.624(2) & 18.623 &  18.960(3)\phantom{1} & 18.959 &  20.194(5)\phantom{1}   & 20.180 \\
WDFS0639-57 &  17.322(3)\phantom{1} & 17.325 &  17.653(4)\phantom{1} & 17.655 &  18.178(3) & 18.177 &  18.639(3) & 18.638 &  19.017(2)\phantom{1} & 19.017 &  20.380(6)\phantom{1}   & 20.367 \\
WDFS0727+32 &  17.164(3)\phantom{1} & 17.167 &  17.471(3)\phantom{1} & 17.474 &  17.993(3) & 17.992 &  18.457(2) & 18.456 &  18.837(3)\phantom{1} & 18.837 &  20.217(7)\phantom{1}   & 20.203 \\
WDFS0815+07 &  18.950(6)\phantom{1} & 18.954 &  19.264(8)\phantom{1} & 19.266 &  19.716(5) & 19.715 &  20.184(5) & 20.183 &  20.579(6)\phantom{1} & 20.579 &  21.962(24)   & 21.945 \\
WDFS0956-38 &  17.698(3)\phantom{1} & 17.701 &  17.859(3)\phantom{1} & 17.862 &  17.862(3) & 17.861 &  18.179(2) & 18.178 &  18.497(2)\phantom{1} & 18.496 &  19.690(5)\phantom{1}   & 19.678 \\
WDFS1024-00 &  18.261(18) & 18.264 &  18.514(4)\phantom{1} & 18.517 &  18.904(5) & 18.903 &  19.317(4) & 19.316 &  19.665(10) & 19.665 &  20.991(13)   & 20.976 \\
WDFS1055-36 &  17.370(2)\phantom{1} & 17.374 &  17.653(2)\phantom{1} & 17.656 &  18.013(2) & 18.012 &  18.427(2) & 18.426 &  18.793(3)\phantom{1} & 18.793 &  20.135(5)\phantom{1}   & 20.122 \\
WDFS1110-17 &  17.041(3)\phantom{1} & 17.044 &  17.354(4)\phantom{1} & 17.357 &  17.867(3) & 17.866 &  18.314(2) & 18.312 &  18.689(2)\phantom{1} & 18.688 &  20.057(5)\phantom{1}   & 20.044 \\
WDFS1111+39 &  17.443(4)\phantom{1} & 17.446 &  17.830(6)\phantom{1} & 17.832 &  18.421(3) & 18.420 &  18.939(4) & 18.938 &  19.344(3)\phantom{1} & 19.344 &  20.797(9)\phantom{1}   & 20.783 \\
WDFS1206+02 &  18.240(4)\phantom{1} & 18.243 &  18.489(4)\phantom{1} & 18.491 &  18.672(4) & 18.671 &  19.060(3) & 19.059 &  19.411(7)\phantom{1} & 19.411 &  20.703(9)\phantom{1}   & 20.689 \\
WDFS1206-27 &  15.737(3)\phantom{1} & 15.740 &  16.041(2)\phantom{1} & 16.043 &  16.476(2) & 16.475 &  16.923(3) & 16.922 &  17.293(2)\phantom{1} & 17.293 &  18.649(4)\phantom{1}   & 18.638 \\
WDFS1214+45 &  16.940(2)\phantom{1} & 16.944 &  17.283(2)\phantom{1} & 17.285 &  17.761(2) & 17.760 &  18.236(3) & 18.235 &  18.629(2)\phantom{1} & 18.629 &  20.038(4)\phantom{1}   & 20.025 \\
WDFS1302+10 &  16.188(2)\phantom{1} & 16.192 &  16.522(2)\phantom{1} & 16.524 &  17.036(2) & 17.036 &  17.514(2) & 17.513 &  17.904(2)\phantom{1} & 17.904 &  19.303(4)\phantom{1}   & 19.292 \\
WDFS1314-03 &  18.258(4)\phantom{1} & 18.261 &  18.597(5)\phantom{1} & 18.599 &  19.102(5) & 19.101 &  19.567(5) & 19.566 &  19.955(9)\phantom{1} & 19.955 &  21.328(12)   & 21.313 \\
WDFS1434-28 &  17.838(4)\phantom{1} & 17.842 &  17.977(4)\phantom{1} & 17.979 &  17.968(3) & 17.967 &  18.285(2) & 18.284 &  18.584(2)\phantom{1} & 18.584 &  19.759(5)\phantom{1}   & 19.747 \\
WDFS1514+00 &  15.110(2)\phantom{1} & 15.114 &  15.391(2)\phantom{1} & 15.393 &  15.709(2) & 15.708 &  16.120(2) & 16.119 &  16.471(1)\phantom{1} & 16.471 &  17.787(4)\phantom{1}   & 17.778 \\
WDFS1535-77 &  15.599(3)\phantom{1} & 15.603 &  15.969(2)\phantom{1} & 15.971 &  16.553(2) & 16.552 &  17.050(2) & 17.048 &  17.457(1)\phantom{1} & 17.457 &  18.890(3)\phantom{1}   & 18.879 \\
WDFS1557+55 &  16.500(2)\phantom{1} & 16.504 &  16.877(2)\phantom{1} & 16.879 &  17.470(3) & 17.469 &  17.992(2) & 17.991 &  18.388(2)\phantom{1} & 18.388 &  19.834(5)\phantom{1}   & 19.822 \\
WDFS1638+00 &  18.016(8)\phantom{1} & 18.019 &  18.318(4)\phantom{1} & 18.320 &  18.840(5) & 18.839 &  19.281(3) & 19.280 &  19.660(5)\phantom{1} & 19.660 &  20.996(9)\phantom{1}   & 20.982 \\
WDFS1814+78 &  15.791(2)\phantom{1} & 15.795 &  16.121(2)\phantom{1} & 16.124 &  16.544(2) & 16.543 &  17.006(2) & 17.004 &  17.393(1)\phantom{1} & 17.392 &  18.786(2)\phantom{1}   & 18.775 \\
WDFS1837-70 &  17.642(3)\phantom{1} & 17.646 &  17.791(3)\phantom{1} & 17.794 &  17.770(2) & 17.770 &  18.092(2) & 18.091 &  18.411(2)\phantom{1} & 18.411 &  19.606(5)\phantom{1}   & 19.594 \\
WDFS1930-52 &  16.729(2)\phantom{1} & 16.733 &  17.034(2)\phantom{1} & 17.036 &  17.484(2) & 17.483 &  17.927(2) & 17.926 &  18.301(2)\phantom{1} & 18.300 &  19.655(5)\phantom{1}   & 19.643 \\
WDFS2101-05 &  18.068(4)\phantom{1} & 18.072 &  18.334(4)\phantom{1} & 18.337 &  18.656(3) & 18.655 &  19.064(2) & 19.063 &  19.414(4)\phantom{1} & 19.414 &  20.740(8)\phantom{1}   & 20.726 \\
WDFS2317-29 &  17.897(3)\phantom{1} & 17.900 &  18.141(3)\phantom{1} & 18.143 &  18.349(3) & 18.348 &  18.748(3) & 18.746 &  19.106(3)\phantom{1} & 19.105 &  20.423(6)\phantom{1}   & 20.410 \\
WDFS2329+00 &  17.943(4)\phantom{1} & 17.947 &  18.109(4)\phantom{1} & 18.111 &  18.161(6) & 18.160 &  18.470(3) & 18.469 &  18.775(7)\phantom{1} & 18.775 &  19.995(6)\phantom{1}   & 19.982 \\
WDFS2351+37 &  17.449(4)\phantom{1} & 17.453 &  17.662(3)\phantom{1} & 17.664 &  18.075(3) & 18.074 &  18.459(3) & 18.458 &  18.787(2)\phantom{1} & 18.787 &  20.075(4)\phantom{1}   & 20.062 \\
\enddata
\tablenotetext{a}{Values in parentheses are 1$\sigma$ errors in mmag. Values in cF columns are corrected values (see equation 7).  Note that it is these corrected values which the SEDs integrated over the WFC3 passbands are expected to match.}
\end{deluxetable*}

\section{Analysis}

\subsection{Previous Analysis Procedure}
\label{sec:danaly}
The analysis presented in this paper incorporates the analysis in \citetalias{Narayan19}, which is based in turn on \citetalias{Narayan16}, as an integral part. In particular, we use the same \texttt{Tlusty}~\citep{Hubeny95} v202 NLTE model atmosphere grid\footnote{\url{http://nova.astro.umd.edu/Tlusty2002/tlusty-frames-refs.html}} as \citetalias{Narayan16}. The grid has 31 uneven steps in $T_\text{eff}$ from 16,000--90,000~K, with a spacing of 2,000~K from 16,000--20,000~K and 2,500~K from 20,000--90,000~K. The grid has 6 even steps in $\log g$ from 7--9.5 dex, with 0.5 dex spacing. The grid covers a wavelength range of 1,350~\AA\ -- 2.7~\micron, in 1~\AA\ steps from 1,350~\AA\ $\le \lambda \le$3,000~\AA, 0.5~\AA\ steps from 3,000~\AA\ $\le \lambda \le$~7,000~\AA, and 5~\AA\ steps for $\lambda >$~7,000~\AA.
\citetalias{Narayan16} used the shape of the observed spectrum, particularly the Balmer lines, to derive \teff\ and \logg. Reddening was deduced, and the process iterated. In \citetalias{Narayan19} we solved for the stellar parameters and the reddening simultaneously, while also using the entire spectrum. Uncertainties in flux calibration were taken into account. The output was a set of best values and distribution of errors for \teff, \logg, and A$_V$, assuming that the ratio of total to selective extinction is $R_V$=3.1. No evidence was seen for variation in $R_V$ for these stars, though the data set is not well suited to detect it. \citetalias{Narayan19} contains all the details.

Synthetic magnitudes in this series of papers are AB magnitudes, defined by \citet{Fukugita96}:
\begin{equation}
    m_{AB} = -2.5 \log{ \frac{\int_{0}^{\infty} \nu^{-1} \cdot F(\nu) \cdot R(\nu) \cdot d\nu}{\int_{0}^{\infty} \nu^{-1} \cdot R(\nu) \cdot d\nu } } - 48.60
\end{equation}
where $F(\nu)$ is the energy flux per unit frequency and $R(\nu)$ is the system response function.

The calibration for both the Northern and equatorial DAWDs (Cycle 20 + 22) and for the Southern DAWDs (Cycle 25) is tied to the published flux values for the three primary CALSPEC standards.
Note that our photometry is tied to the previous flux calibration from \citet{Bohlin2014a}, which defined CALSPEC until 2019. A flux calibration based on new models for the three CALSPEC primary standards was released in 2020 \citep{Bohlin2020}; subsequently, an updated time-dependent calibration for the WFC3 UVIS and IR detectors was also delivered in October 2020 \citep{Calamida2022detector}.
The WFC3 UVIS detector has indeed had an average sensitivity decline of $\approx$ 0.15\%/year, differing depending on the filter \citep{Calamida2022detector}. A sensitivity decline of the WFC3 IR detector has not been established yet; however, preliminary evidence indicate an average decline of $\approx$ 0.1\%/year (\citealt{Bohlin19}; Bajaj et al. 2022\footnote{\url{https://www.stsci.edu/files/live/sites/www/files/home/hst/instrumentation/wfc3/ documentation/instrument-science-reports-isrs/\_documents/2022/WFC3-ISR-2022-07.pdf}})

Therefore, we verified our photometry for time sensitivity changes as discussed in detail in \citetalias{Calamida19}. 
Although we applied an offset to bring Cycle 20 photometry onto the Cycle 22 system, we did not measure a sensitivity change in the Cycle 22 photometry of the three CALSPEC primary DAWDs, spanning approximately 1.5 years. We then did not correct for time sensitivity changes the photometry of the Northern and equatorial DAWDs, nor the photometry for the Southern DAWDs, collected during Cycle 25.

The resulting HST photometry for the 32 established DAWDs and our measurements of the 3 CALSPEC standards is listed in Table \ref{tab:HST_Residuals}.

\begin{deluxetable*}{ccccccccccccc}
\tablecaption{HST Synthetic Photometry and Residuals when Compared
  with Measured Values. Units are AB mag.\tablenotemark{a}\label{tab:HST_Residuals}}
\scriptsize
\tablehead{
\colhead{Object} & \colhead{F275W} & \colhead{F275W} & \colhead{F336W}
& \colhead{F336W} & \colhead{F475W} & \colhead{F475W} &
\colhead{F625W} & \colhead{F625W} & \colhead{F775W} & \colhead{F775W}
& \colhead{F160W} & \colhead{F160W} \\
\colhead{} & \colhead{Synth.} & \colhead{Resid.} & \colhead{Synth.} &
\colhead{Resid.} & \colhead{Synth.} & \colhead{Resid.} &
\colhead{Synth.} & \colhead{Resid.} & \colhead{Synth.} &
\colhead{Resid.} & \colhead{Synth.} & \colhead{Resid.}}
\startdata
G191B2B & 10.493 &     \phantom{$-$}0.001 & 10.891 &     \phantom{$-$}0.002 & 11.502 &    $-$0.004 & 12.032 &    $-$0.002 & 12.448 &     \phantom{$-$}0.003 & 13.879 &     0.004 \\
GD153 & 12.206 &    $-$0.001 & 12.569 &     \phantom{$-$}0.001 & 13.095 &     \phantom{$-$}0.004 & 13.596 &     \phantom{$-$}0.001 & 14.002 &    $-$0.000 & 15.414 &    $-$0.005 \\
GD71 & 11.994 &    $-$0.001 & 12.337 &     \phantom{$-$}0.001 & 12.796 &     \phantom{$-$}0.002 & 13.276 &     \phantom{$-$}0.002 & 13.674 &    $-$0.002 & 15.066 &    $-$0.003 \\
WDFS0103-00 & 18.193 &     \phantom{$-$}0.005 & 18.536 &    $-$0.007 & 19.088 &    $-$0.005 & 19.569 &    $-$0.001 & 19.957 &     \phantom{$-$}0.008 & 21.337 &     0.004 \\
WDFS0122-30 & 17.672 &     \phantom{$-$}0.003 & 18.000 &    $-$0.004 & 18.458 &     \phantom{$-$}0.001 & 18.925 &    $-$0.004 & 19.314 &     0.006 & 20.693 &    $-$0.002 \\
WDFS0228-08 & 19.519 &     \phantom{$-$}0.002 & 19.708 &     \phantom{$-$}0.009 & 19.827 &    $-$0.013 & 20.172 &    $-$0.004 & 20.495 &     \phantom{$-$}0.006 & 21.719 &     \phantom{$-$}0.003 \\
WDFS0238-36 & 17.790 &     \phantom{$-$}0.003 & 17.976 &    $-$0.001 & 18.096 &    $-$0.002 & 18.436 &     \phantom{$-$}0.002 & 18.757 &     \phantom{$-$}0.000 & 19.977 &     \phantom{$-$}0.002 \\
WDFS0248+33 & 17.832 &     \phantom{$-$}0.000 & 18.048 &    $-$0.005 & 18.369 &    $-$0.000 & 18.745 &    $-$0.000 & 19.074 &     \phantom{$-$}0.003 & 20.339 &    $-$0.013 \\
WDFS0458-56 & 17.027 &     \phantom{$-$}0.000 & 17.353 &     \phantom{$-$}0.000 & 17.754 &    $-$0.000 & 18.215 &     \phantom{$-$}0.000 & 18.604 &    $-$0.003 & 19.975 &     \phantom{$-$}0.012 \\
WDFS0541-19 & 18.026 &    $-$0.002 & 18.216 &     \phantom{$-$}0.001 & 18.272 &     \phantom{$-$}0.003 & 18.626 &    $-$0.002 & 18.958 &     \phantom{$-$}0.002 & 20.184 &    $-$0.004 \\
WDFS0639-57 & 17.328 &    $-$0.003 & 17.650 &     \phantom{$-$}0.006 & 18.177 &    $-$0.000 & 18.638 &    $-$0.000 & 19.015 &     \phantom{$-$}0.002 & 20.377 &    $-$0.010 \\
WDFS0727+32 & 17.161 &     \phantom{$-$}0.006 & 17.478 &    $-$0.004 & 17.998 &    $-$0.006 & 18.457 &    $-$0.002 & 18.832 &     \phantom{$-$}0.004 & 20.190 &     \phantom{$-$}0.014 \\
WDFS0815+07 & 18.949 &     \phantom{$-$}0.005 & 19.268 &    $-$0.003 & 19.720 &    $-$0.004 & 20.186 &    $-$0.003 & 20.571 &     \phantom{$-$}0.008 & 21.941 &     \phantom{$-$}0.005 \\
WDFS0956-38 & 17.707 &    $-$0.005 & 17.864 &    $-$0.002 & 17.851 &     \phantom{$-$}0.009 & 18.178 &    $-$0.001 & 18.496 &     \phantom{$-$}0.000 & 19.692 &    $-$0.014 \\
WDFS1024-00 & 18.258 &     \phantom{$-$}0.006 & 18.513 &     \phantom{$-$}0.004 & 18.911 &    $-$0.008 & 19.315 &     \phantom{$-$}0.001 & 19.663 &     \phantom{$-$}0.002 & 20.969 &     \phantom{$-$}0.007 \\
WDFS1055-36 & 17.374 &    $-$0.000 & 17.658 &    $-$0.002 & 18.008 &     \phantom{$-$}0.004 & 18.429 &    $-$0.003 & 18.794 &    $-$0.001 & 20.121 &     \phantom{$-$}0.001 \\
WDFS1110-17 & 17.046 &    $-$0.002 & 17.358 &    $-$0.001 & 17.862 &     \phantom{$-$}0.004 & 18.314 &    $-$0.001 & 18.688 &     \phantom{$-$}0.001 & 20.046 &    $-$0.002 \\
WDFS1111+39 & 17.446 &     \phantom{$-$}0.000 & 17.830 &     \phantom{$-$}0.003 & 18.420 &    $-$0.001 & 18.936 &     \phantom{$-$}0.002 & 19.346 &    $-$0.002 & 20.767 &     \phantom{$-$}0.016 \\
WDFS1206-27 & 15.741 &    $-$0.001 & 16.043 &     \phantom{$-$}0.000 & 16.476 &    $-$0.001 & 16.918 &     \phantom{$-$}0.004 & 17.292 &     \phantom{$-$}0.001 & 18.646 &    $-$0.007 \\
WDFS1206+02 & 18.246 &    $-$0.002 & 18.486 &     \phantom{$-$}0.005 & 18.673 &    $-$0.002 & 19.060 &    $-$0.001 & 19.410 &     \phantom{$-$}0.001 & 20.685 &     \phantom{$-$}0.004 \\
WDFS1214+45 & 16.944 &    $-$0.001 & 17.285 &    $-$0.000 & 17.758 &     \phantom{$-$}0.002 & 18.236 &    $-$0.000 & 18.631 &    $-$0.002 & 20.022 &     \phantom{$-$}0.003 \\
WDFS1302+10 & 16.189 &     \phantom{$-$}0.002 & 16.525 &    $-$0.002 & 17.039 &    $-$0.003 & 17.513 &    $-$0.000 & 17.903 &     \phantom{$-$}0.001 & 19.288 &     \phantom{$-$}0.004 \\
WDFS1314-03 & 18.266 &    $-$0.004 & 18.592 &     \phantom{$-$}0.007 & 19.100 &     \phantom{$-$}0.001 & 19.567 &    $-$0.001 & 19.951 &     \phantom{$-$}0.004 & 21.325 &    $-$0.011 \\
WDFS1434-28 & 17.841 &     \phantom{$-$}0.000 & 17.977 &     \phantom{$-$}0.002 & 17.974 &    $-$0.007 & 18.281 &     \phantom{$-$}0.004 & 18.583 &     \phantom{$-$}0.001 & 19.756 &    $-$0.009 \\
WDFS1514+00 & 15.119 &    $-$0.005 & 15.389 &     \phantom{$-$}0.004 & 15.707 &     \phantom{$-$}0.001 & 16.115 &     \phantom{$-$}0.004 & 16.474 &    $-$0.003 & 17.785 &    $-$0.007 \\
WDFS1535-77 & 15.598 &     \phantom{$-$}0.004 & 15.971 &    $-$0.000 & 16.555 &    $-$0.003 & 17.052 &    $-$0.004 & 17.456 &     \phantom{$-$}0.001 & 18.875 &     \phantom{$-$}0.004 \\
WDFS1557+55 & 16.502 &     \phantom{$-$}0.002 & 16.882 &    $-$0.003 & 17.470 &    $-$0.001 & 17.986 &     \phantom{$-$}0.005 & 18.394 &    $-$0.006 & 19.810 &     \phantom{$-$}0.012 \\
WDFS1638+00 & 18.017 &     \phantom{$-$}0.003 & 18.322 &    $-$0.002 & 18.836 &     \phantom{$-$}0.003 & 19.283 &    $-$0.004 & 19.650 &     \phantom{$-$}0.010 & 20.992 &    $-$0.011 \\
WDFS1814+78 & 15.795 &    $-$0.000 & 16.123 &     \phantom{$-$}0.001 & 16.542 &     \phantom{$-$}0.002 & 17.005 &    $-$0.001 & 17.395 &    $-$0.002 & 18.769 &     \phantom{$-$}0.006 \\
WDFS1837-70 & 17.643 &     \phantom{$-$}0.003 & 17.793 &     \phantom{$-$}0.000 & 17.772 &    $-$0.003 & 18.093 &    $-$0.003 & 18.407 &     \phantom{$-$}0.004 & 19.596 &    $-$0.002 \\
WDFS1930-52 & 16.735 &    $-$0.002 & 17.035 &     \phantom{$-$}0.001 & 17.482 &     \phantom{$-$}0.001 & 17.927 &    $-$0.001 & 18.300 &     \phantom{$-$}0.001 & 19.651 &    $-$0.007 \\
WDFS2101-05 & 18.073 &    $-$0.001 & 18.336 &     \phantom{$-$}0.001 & 18.655 &    $-$0.000 & 19.062 &    \phantom{$-$} 0.010 & 19.417 &    $-$0.003 & 20.723 &     \phantom{$-$}0.003 \\
WDFS2317-29 & 17.898 &     \phantom{$-$}0.002 & 18.154 &    $-$0.011 & 18.344 &     \phantom{$-$}0.004 & 18.747 &    $-$0.000 & 19.107 &    $-$0.001 & 20.397 &     \phantom{$-$}0.013 \\
WDFS2329+00 & 17.949 &    $-$0.003 & 18.111 &     \phantom{$-$}0.000 & 18.146 &     \phantom{$-$}0.014 & 18.470 &    $-$0.001 & 18.784 &    $-$0.009 & 19.982 &    $-$0.000 \\
WDFS2351+37 & 17.445 &     \phantom{$-$}0.008 & 17.673 &    $-$0.009 & 18.074 &     \phantom{$-$}0.000 & 18.455 &     \phantom{$-$}0.003 & 18.786 &     \phantom{$-$}0.000 & 20.064 &    $-$0.002 \\
\enddata
\tablenotetext{a}{Residual value is observed photometry minus synthetic photometry.}
\end{deluxetable*}

\subsection{Analysis Overview}
\label{sec:analys-a}
\citetalias{Narayan19} argued for better solutions to the spectroscopic and photometric parameters by doing a complete hierarchical Bayesian
model (e.g., \citet{Loredo19}), solving for all stars (both the stars presented in \citetalias{Narayan19} and the new stars presented here) simultaneously. The analysis presented here takes a significant step in this direction. We go into detail below, but first lay out the general idea.

The new analysis draws on the lessons learned from \citetalias{Narayan16} and \citetalias{Narayan19} and attempts to make incremental improvements. Our goals are to
\begin{itemize}
\item Perform a full Bayesian analysis incorporating the spectroscopy and photometry for all DAWDs simultaneously.
\item Preserve the alternative analysis of \citetalias{Narayan16} which removed the dependence of the results on the MAST zeropoints for the CALSPEC primary standards.
\item Account for the count rate nonlinearity (CRNL) of the F160W data through free model parameters.
\end{itemize}

An analysis based on \citetalias{Narayan19} for all the stars observed in Cycle 20, 22, and 25 provides input priors and error distributions for the spectroscopic parameters
and HST photometry. 
 
The input HST photometry can be in any magnitude system that is stable over time, including instrumental, as was used in \citetalias{Narayan16}.  In the process of matching the observed photometry to the synthetic photometry from the DAWD models, all color dependent offsets in the input magnitude system relative to an AB system are corrected, leaving only an overall absolute flux calibration to determine. In the results presented here, the incoming magnitudes have been initially placed on the CALSPEC system using the procedure in \citetalias{Narayan19}. The absolute flux calibration is not altered from its CALSPEC value.
The stellar parameters, \teff, \logg, \av\ and distance modulus, are allowed to change independently for each star, and the per-band zeropoints are allowed to vary while keeping the overall flux normalization fixed.

The input photometry includes our observations of the CALSPEC standards as well as our observations of the program stars. 

The errors in this new analysis method should be smaller than the errors in the \citetalias{Narayan19} method, simply because band-to-band differences effectively take out small errors in CALSPEC.  This expectation is borne out, as shown in Figure \ref{fig:WDmodelPhotResid}.

\subsection {Analysis Details}
\label{sec:analys-b}
The first goal is limited by available computer power.  Determining the posterior distribution for a model which accounts for the spectral and photometric data from all DAWDs simultaneously is judged to be impractical currently.  Recognizing that the determination of \teff\ and \logg\ relies almost exclusively on the spectroscopy and is nearly independent of the HST photometry, while the determination of Av and the distance modulus of each DAWD are nearly independent of the spectroscopy, we settled on a practical compromise with the following outline:
\begin{enumerate}
\item The analysis of \citetalias{Narayan19} is performed as before for each DAWD separately.  This yields for each DAWD, $s$, the posterior distribution for the apparent magnitudes in band $\lambda$, $m_{s}^{\lambda}$, and the SED parameters \teff, \logg, and \av\
\item Using the posteriors from the previous step as input priors, the photometry of all DAWDs are incorporated simultaneously in a Bayesian model, the posteriors of which yield a second determination of the \teff, \logg, and \av\ posteriors, together with those for the per-band zeropoint shifts $\delta_{\lambda}$ and the F160W CRNL slope, $\alpha_{CRNL}$.
\end{enumerate}

If necessary for convergence, the two steps of this calculation could be iterated, incorporating the zeropoint shifts and CRNL slope from step 2 as priors into step 1.  We have determined that this iteration is not necessary, a conclusion which supports the assumption of very weak coupling between the modeling of the spectroscopy and the photometry.

The calculation proceeds as follows
\begin{enumerate}
\item For each DAWD, a 2D normal distribution is fit to the output \teff, \logg\ chain from the \citetalias{Narayan19} analysis (preceding section).  These are used as priors.

\item Noninformative priors are used for \av, with the exception of those for the primary CALSPEC DAWDs. 
The \av\ values of the three primary standards are constrained with an upper limit of 0.003, consistent with CALSPEC upper limits \citep{Bohlin2020}.  This is a crucial element of the calculation and is the only way that the CALSPEC DAWDs play a special role.
\item The likelihood function is constructed, utilizing the same synthetic spectral model grid employed in \citetalias{Narayan19}.\item A set of MCMC chains is run using emcee (\citealt{Foreman-Mackey13}).
\item The posterior distributions are constructed for the output chains for the per-object \teff, \logg, \av, the overall model per-band zeropoint shifts, $\Delta_{\lambda}$, and the F160W CRNL slope, $\alpha_{CRNL}$.
\item As a consistency check, the \teff, \logg, and \av\ posteriors are compared with those from the separate \citetalias{Narayan19} DAWD analysis performed above.  Major differences would be cause for further investigation, but in practice have not been found.
\end{enumerate}

\subsubsection{Likelihood function}
The likelihood function for each DAWD is a small modification of that employed in \citetalias{Narayan19}.
\begin{equation}
    \begin{split}
        P(\{m_s\} & |\, T_{\rm eff}, \log g, A_V, R_V, \mu_s, \Delta_{\lambda}) = \\
        & \prod_{\lambda = 1}^{N_{\text{PB}}}\, N(m_{s,\lambda}  |\, M_{s,\lambda}(T_{\rm eff}, \log g, A_V, R_V) + \mu_s + \Delta_{\lambda}, \sigma_{s,\lambda}) \\
    \end{split}
\end{equation}
where $m_{s,\lambda}$ is the observed magnitude for a DAWD star $s$ in \emph{HST/WFC3} passband $\lambda \in \{$\textit{F275W, F336W, F475W, F625W, F775W, F160W}$\}$, with photometric measurement error described by an estimated standard deviation $\sigma_{s\lambda}$, $M_{s,\lambda}(T_{\rm eff}, \log g, A_V, R_V)$ is the synthetic magnitude of the reddened SED through passband $\lambda$, and $N(m, \sigma)$ is the normal distribution. 
$\mu_s$ is a per-star achromatic normalization parameter which is added to the synthetic reddened magnitudes in all passbands to account for the distance and radius of the DAWD $s$. $\Delta_{\lambda}$ was introduced in \citetalias{Narayan16} and is the \emph{star independent} offset to the observed magnitudes in passband $\lambda$ to convert them to AB magnitudes (the magnitude system for the synthetic magnitudes).  If the $\Delta_{\lambda}$ were left unconstrained, there would be a degeneracy between their mean value and the mean value of the $\mu_s$ over the full set of stars.  To break the degeneracy, we require

\begin{equation}
    \sum_{\lambda = 1}^{N_{\text{PB}}}\, \Delta_{\lambda} = 0
\end{equation}
The observed magnitudes for this analysis are already on the CALSPEC AB system, and the $\Delta_{\lambda}$ are expected to be quite small, accounting only for small errors in the measured HST passbands and/or aperture corrections, an expectation that is realized, as shown in Table \ref{tab:zp}.  However, this constraint is not the only possible choice for breaking the degeneracy.  One could, for example, instead require that the synthetic SED of a selected calibration star at a given wavelength match a value determined outside the system (e.g., CALSPEC).

\begin{deluxetable*}{ccc}
\tablecaption{Values for $\Delta_\lambda$ and Their Uncertainties\label{tab:zp}}
\tablehead{
  \colhead{Band} & \colhead{$\Delta$} & \colhead{$\Delta_{\sigma}$}\\
\colhead{} & \colhead{mag}& \colhead{mag}}
\startdata
F275W &   $-$0.004 &    0.001 \\
F336W &   $-$0.002 &    0.001 \\
F475W &    \phantom{$-$}0.001 &    0.000 \\
F625W &    \phantom{$-$}0.001 &    0.001 \\
F775W &    \phantom{$-$}0.000 &    0.001 \\
F160W &    \phantom{$-$}0.004 &    0.001 \\
\enddata
\end{deluxetable*}

The likelihood function for the entire model is then the product over all DAWDs of the likelihood for each individual DAWD. For the results reported here, $R_V=3.1$.

There is one further refinement beyond the previous analysis.  The HST detector for the F160W band is known to have a dependence of the counts from a source integrated over the exposure time on the \emph{rate} of those counts, commonly referred to as the ``count rate nonlinearity'' (CRNL) \citep{Bohlin19, Riessetal2019}. 
To account for this effect, a synthetic magnitude in F160W $M_{s, F160W}$, is observed as
\begin{equation}
    M_{s, F160W} + \alpha_{CRNL} (M_{s, F160W} - \beta_{CRNL})
\end{equation}

We include $\alpha_{CRNL}$ in the free parameters of the model. We have found no significant effects from varying $\beta_{CRNL}$, and it is arbitrarily fixed at 15. The $\lambda = F160W$ term in the product in Equation 2 then becomes:

\begin{equation}
    \begin{split}
        N(m_{s,F160W} | &( M_{s,F160W} + \\
        &\alpha_{CRNL}  (M_{s,F160W}-  \beta_{CRNL}) + \Delta_{F160W}+ \mu_s , \sigma_{s,F160W})) \\
    \end{split}
\end{equation}

It is convenient to express equation 2 as

\begin{equation}
    \begin{split}
        P(\{m_s\} & |\, T_{\rm eff}, \log g, A_V, R_V, \mu_s, \Delta_{\lambda}) = \\
        & \prod_{\lambda = 1}^{N_{\text{PB}}}\, N(m_{s,\lambda}^{corr}(\Delta_{\lambda})|\, M_{s,\lambda}(T_{\rm eff}, \log g, A_V, R_V) + \mu_s, \sigma_{s,\lambda}) \\
    \end{split}
\end{equation}

where the observed magnitude, corrected for the zero point shift, and in the case of F160W, the CRNL, is
\begin{equation}
m^{corr}_{s,\lambda} = m_{s,\lambda} - \Delta_{\lambda} + \delta_{\lambda, F160W} \alpha_{CRNL} (m_{s,F160W} - \beta_{CRNL})
\end{equation}

The values of $m^{corr}_{s,\lambda}$ are given in Table \ref{tab:HST_Residuals} alongside those for $m_{s,\lambda}$. Note that it is these values which the SEDs integrated over the WFC3 passbands are expected to match.

The free model parameters, then, include \teff, \logg, \av, and $\mu_s$ for each star,  the five element array $\Delta_{\lambda}$, and $\alpha_{CRNL}$, a total of 146.

\section{Results}
\label{sec:results}
We used the emcee implementation of MCMC to sample the posterior probability density function (pdf) of the model parameters employing 400 walkers, each producing a chain of 20000 steps after a 100 step burn-in.  Corner plots of the resulting pdf show good convergence in all parameters.  Two examples are shown in Figures \ref{fig:corner} and \ref{fig:cornerCRNL}.  Derived parameters for each DAWD are presented in Table \ref{tab:Object_parameters}.

\begin{deluxetable*}{ccccccc}
  \scriptsize
 \tablecaption{Derived Object Parameters\tablenotemark{a}\label{tab:Object_parameters}}
 \tablehead{
\colhead{Object} & \colhead{\teff (K)} & \colhead{\teff$\sigma$ (K)} & \colhead{$\log~g$ (dex)}& \colhead{$\log~g$ $\sigma$ (dex)} & \colhead{\av\ (mag)}   & \colhead{\av\ $\sigma$ (mag)}}
\startdata
G191B2B     & 63200 &  447&    7.588 &    0.032 &    0.001  &    0.001 \\
GD153       & 38765 &  185&    7.720 &    0.036 &    0.001  &    0.001 \\
GD71        & 32705 &   90&    7.782 &    0.020 &    0.003  &    0.001 \\
WDFS0103-00 & 57959 & 2366&    7.678 &    0.081 &    0.119  &    0.008 \\
WDFS0122-30 & 33964 &  215&    7.771 &    0.031 &    0.048  &    0.005 \\
WDFS0228-08 & 23026 &  269&    7.831 &    0.041 &    0.156  &    0.014 \\
WDFS0238-36 & 23169 &   84&    7.880 &    0.014 &    0.171  &    0.005 \\
WDFS0248+33 & 33148 &  393&    7.103 &    0.043 &    0.305  &    0.007 \\
WDFS0458-56 & 30111 &   78&    7.788 &    0.018 &    0.014  &    0.003 \\
WDFS0541-19 & 20436 &   83&    7.829 &    0.014 &    0.053  &    0.006 \\
WDFS0639-57 & 54760 &  890&    7.898 &    0.048 &    0.162  &    0.004 \\
WDFS0727+32 & 53516 & 1364&    7.697 &    0.064 &    0.167  &    0.005 \\
WDFS0815+07 & 35008 &  758&    7.297 &    0.049 &    0.076  &    0.012 \\
WDFS0956-38 & 19219 &   63&    7.875 &    0.012 &    0.078  &    0.005 \\
WDFS1024-00 & 36021 &  959&    7.654 &    0.125 &    0.240  &    0.015 \\
WDFS1055-36 & 29503 &  103&    7.930 &    0.025 &    0.106  &    0.005 \\
WDFS1110-17 & 46442 & 1014&    8.011 &    0.080 &    0.159  &    0.005 \\
WDFS1111+39 & 56874 & 1226&    7.799 &    0.041 &    0.022  &    0.005 \\
WDFS1206+02 & 23647 &  203&    7.886 &    0.021 &    0.056  &    0.011 \\
WDFS1206-27 & 33884 &  169&    7.901 &    0.033 &    0.111  &    0.004 \\
WDFS1214+45 & 34169 &  255&    7.846 &    0.038 &    0.022  &    0.005 \\
WDFS1302+10 & 41577 &  634&    7.927 &    0.017 &    0.080  &    0.005 \\
WDFS1314-03 & 43200 & 1397&    7.823 &    0.091 &    0.110  &    0.010 \\
WDFS1434-28 & 20332 &   86&    7.818 &    0.016 &    0.177  &    0.005 \\
WDFS1514+00 & 28576 &  127&    7.903 &    0.013 &    0.120  &    0.005 \\
WDFS1535-77 & 50524 &  806&    9.080 &    0.029 &    0.034  &    0.004 \\
WDFS1557+55 & 57758 &  983&    7.551 &    0.070 &    0.029  &    0.004 \\
WDFS1638+00 & 58415 & 2133&    7.749 &    0.108 &    0.210  &    0.008 \\
WDFS1814+78 & 31048 &  130&    7.802 &    0.014 &    0.021  &    0.004 \\
WDFS1837-70 & 19199 &   63&    7.869 &    0.012 &    0.094  &    0.005 \\
WDFS1930-52 & 36263 &  191&    7.669 &    0.020 &    0.132  &    0.003 \\
WDFS2101-05 & 29187 &  239&    7.766 &    0.026 &    0.145  &    0.009 \\
WDFS2317-29 & 23120 &   48&    7.851 &    0.019 &    0.001  &    0.002 \\
WDFS2329+00 & 20557 &  196&    7.957 &    0.030 &    0.129  &    0.011 \\
WDFS2351+37 & 41208 &  842&    7.702 &    0.081 &    0.332  &    0.007 \\
\enddata
\tablenotetext{a}{Parameter values are the medians of the posterior distributions.}
\end{deluxetable*}

\subsection{Comparison with CALSPEC}
Despite the input photometry being on the 2014 CALSPEC system, there are small differences between our model SEDs for the CALSPEC primary DAWDs and the 2014 CALSPEC model SEDs of \citet{Bohlin2014a}.  
The absolute flux scale reported herein and our
previous (\citetalias{Narayan19}) paper is based on the absolute flux calibration of the
WFC3 filters described in \citetalias{Calamida19}. These spectral energy
distributions (SEDs), i.e., absolute flux in physical units, are based on models
of the three primary DAWDs, G191B2B, GD153, and GD71 \citep{Bohlin2014a}. However, those models were improved with new NLTE grids computed by
Ivan Hubeny and Thomas Rauch \citep{Bohlin2020}, which resulted
in changes to the basis of the HST flux scale by up to ~3\% at some wavelengths.
A future paper will report the SEDs of our DAWD standards, as adjusted to the
more recent \citet{Bohlin2020} flux scale. See the Conclusion Section for more details.

\subsection{Comparison with Gentile Fusillo 2021}
    A new catalog of white dwarfs based on Gaia EDR3 was recently published by Gentile Fusillo et al (\citet[hereafter GF21]{Gentile-Fusillo21}). This catalog contains values for the stellar parameters \teff \, and \logg \, based on WD model atmospheres in conjunction with Gaia photometry, and \av \, from a three dimensional extinction model.  It is useful to compare the GF21 values for our WDFS stars with our results.  These are shown in Figure\ref{fig:GF21compare}. The comparison for \av \, is particularly useful, showing good agreement between values determined by two completely independent methods.  The \teff \, and \logg \, comparisons are likewise based on independent methods, but the a priori confidence in the GF21 values must be lowered by the lack of spectroscopic input.

\subsection{Count Rate Nonlinearity}
The value determined for $\alpha_{CRNL}$ is $-1.74 \pm 0.32$ mmag per mag. This is significantly less than the published value of $-3.12 \pm 0.32$ mmag/mag in \citet{Bohlin19}, or the combined result of $-3.0 \pm 0.24$ mmag/mag in \citet{Riessetal2019}. However, the CRNL is consistent with the value of $-2.36 \pm 0.48$ mmag/mag\ for the subset of our stars analyzed in \citet{Riessetal2019}. As shown in Figure \ref{fig:cornerCRNL}, the posterior distribution for $\alpha_{CRNL}$ is tightly constrained.

\subsection{Synthetic Magnitudes for Common Survey Passbands}
\label{sec:surveymags}
As in \citetalias{Narayan19}, we have calculated the synthetic magnitudes for our standards in a number of common survey passbands.  The filter passbands are obtained from the Spanish Virtual Observatory (SVO) Filter Profile Service \citep{SVO20,SVO12}.  To calculate the synthetic magnitudes, we utilize the full MCMC chains from our analysis run, and for each point on the chain calculate the associated synthetic magnitudes. This gives a probability density function for each magnitude, which we characterized by its median and standard deviation.  The standard deviations are typically less than one milli-mag (0.001 mag), significantly less in most cases than the survey reported observational uncertainties, and certainly less than the (unknown) systematic errors.  The standard deviations therefore do not reflect the real uncertainties in our synthetic magnitudes, particularly in passbands not closely aligned to the HST passbands, and we do not include these values in the tables below.

For each survey system, we include below a table of our synthetic magnitudes, and where available the magnitudes and uncertainties reported by the survey.  A plot for each band of each photometric system shows the magnitude differences (in the sense synthetic - observed) as a function of magnitude.  We note some caveats.  The observed magnitudes for each star are derived from differing photometric systems, especially the broad Gaia $G$ filter.  That these independent photometric systems demonstrate good agreement between their observed and our synthetic magnitudes is further evidence for the robustness of our system.

We provide magnitudes for our stars in the DES, DECaLS,
PaNSTARRS1, SDSS, and Gaia systems.

DES observed point-spread-function (PSF) magnitudes (\texttt{WAVG\_MAG\_PSF}) are from DES DR2~\citep{desdr2_2021}. Results are in Table~\ref{tab:DECAM_Residuals}, and in Figure~\ref{fig:DECAMResid}.

\begin{deluxetable*}{ccccccccccccc}
  \scriptsize
  \tablecaption{Observed and Synthetic Photometry in DECam Bands. Units are AB mag.\label{tab:DECAM_Residuals}}
\tablehead{
\colhead{Object}   &  \colhead{$u$}      &  \colhead{$u$}      &\colhead{$g$}   &  \colhead{$g$} & \colhead{$r$}
&  \colhead{$r$}   &  \colhead{$i$}   &  \colhead{$i$}   &  \colhead{$z$}
&  \colhead{$z$}   &  \colhead{$y$}   &  \colhead{$y$} \\
\colhead{}   &  \colhead{Obs.}   &  \colhead{Synth.} &   \colhead{Obs.}
&  \colhead{Synth.}   &  \colhead{Obs.}   &  \colhead{Synth.}   &  \colhead{Obs.} &  \colhead{Synth.}   &  \colhead{Obs.}
&  \colhead{Synth.}   &  \colhead{Obs.}   &  \colhead{Synth.}}
\startdata
G191B2B &     & 11.114 &     &     11.502      &     &  12.092   &     &     12.491   &     &     12.820   &     &     12.982 \\
GD153 &     & 12.772 &     &     13.096      &     &  13.655   &     &     14.044   &     &     14.368   &     &     14.528 \\
GD71 &     & 12.526 &     &     12.797      &     &  13.335   &     &     13.715   &     &     14.035   &     &     14.192 \\
WDFS0103-00 &  & 18.737 &  19.107 (2) &  19.088 &  19.632 (3) &  19.624 &  20.003 (4) &  19.997 &  20.307 (11) &  20.309 &  20.413 (48) &  20.464 \\
WDFS0122-30 &   & 18.185 &  18.471 (1) &  18.458 &  18.978 (1) &  18.982 &  19.346 (2) &  19.355 &  19.661 (5)\phantom{1} &  19.669 &  19.818 (24) &  19.825 \\
WDFS0228-08 &   & 19.801 &  19.816 (3) &  19.827 &  20.218 (3) &  20.221 &  20.527 (6) &  20.531 &  20.816 (14) &  20.802 &  20.997 (73) &  20.936 \\
WDFS0238-36 &  & 18.069 &  18.091 (1) &  18.095 &  18.481 (1) &  18.485 &  18.794 (2) &  18.792 &  19.043 (3)\phantom{1} &  19.063 &  19.219 (12) &  19.196 \\
WDFS0248+33 &     & 18.179 &     &     18.370  &     &     18.792   &     &     19.110   &     &     19.385   &     &     19.523 \\
WDFS0458-56 &   & 17.530 &  17.770 (1) &  17.754 &  18.272 (1) &  18.273 &  18.637 (2) &  18.645 &  18.939 (3)\phantom{1} &  18.959 &  19.134 (15) &  19.113 \\
WDFS0541-19 &   & 18.296 &  18.280 (1) &  18.272 &  18.684 (1) &  18.677 &  18.995 (2) &  18.994 &  19.261 (4)\phantom{1} &  19.271 &  19.410 (15) &  19.404 \\
WDFS0639-57 &     & 17.841 &     &     18.177 &      &     18.692   &     &     19.054   &     &     19.360   &     &     19.512 \\
WDFS0727+32 &     & 17.667 &     &     17.999  &     &     18.511   &     &     18.872   &     &     19.176   &     &     19.328 \\ 
WDFS0815+07 &     & 19.449 &     &     19.721  &     &     20.242   &     &     20.611   &     &     20.921   &     &     21.075 \\
WDFS0956-38 &     & 17.922 &     &     17.851  &     &     18.228   &     &     18.532   &     &     18.799   &     &     18.928 \\
WDFS1024-00 &     & 18.669 &     &     18.911  &     &     19.365   &     &     19.700   &     &     19.989   &     &     20.133 \\
WDFS1055-36 &     & 17.817 &     &     18.008  &     &     18.483   &     &     18.833   &     &     19.133   &     &     19.281 \\
WDFS1110-17 &     & 17.544 &     &     17.861  &     &     18.368   &     &     18.727   &     &     19.032   &     &     19.184 \\
WDFS1111+39 &     & 18.046 &     &     18.420  &     &     18.995   &     &     19.389   &     &     19.715   &     &     19.875 \\
WDFS1206+02 &     & 18.604 &     &     18.673  &     &     19.114   &     &     19.448   &     &     19.737   &     &     19.878 \\
WDFS1206-27 &     & 16.218 &     &     16.476  &     &     16.973   &     &     17.332   &     &     17.638   &     &     17.789 \\
WDFS1214+45 &     & 17.475 &     &     17.759  &     &     18.294   &     &     18.672   &     &     18.990   &     &     19.147 \\
WDFS1302+10 &     & 16.719 &     &     17.039  &     &     17.570   &     &     17.944   &     &     18.258   &     &     18.414 \\
WDFS1314-03 &     & 18.782 &     &     19.100  &     &     19.622   &     &     19.991   &     &     20.302   &     &     20.456 \\
WDFS1434-28 &     & 18.032 &     &     17.973  &     &     18.328   &     &     18.617   &     &     18.875   &     &     19.001 \\
WDFS1514+00 &     & 15.539 &     &     15.707  &     &     16.169   &     &     16.512   &     &     16.808   &     &     16.954 \\
WDFS1535-77 &     & 16.183 &     &     16.553  &     &     17.112   &     &     17.498   &     &     17.823   &     &     17.984 \\
WDFS1557+55 &     & 17.098 &     &     17.470  &     &     18.044   &     &     18.436   &     &     18.761   &     &     18.921 \\
WDFS1638+00 &     & 18.507 &     &     18.837  &     &     19.336   &     &     19.688   &     &     19.988   &     &     20.138 \\
WDFS1814+78 &     & 16.303 &     &     16.542  &     &     17.063   &     &     17.436   &     &     17.750   &     &     17.905 \\
WDFS1837-70 &     & 17.848 &     &     17.771  &     &     18.142   &     &     18.442   &     &     18.708   &     &     18.836 \\
WDFS1930-52 &     & 17.210 &     &     17.482  &     &     17.981   &     &     18.339   &     &     18.643   &     &     18.794 \\
WDFS2101-05 &     & 18.484 &     &     18.655  &     &     19.114   &     &     19.455   &     &     19.748   &     &     19.893 \\
WDFS2317-29 &     & 18.276 &     &     18.345  &     &     18.802   &     &     19.145   &     &     19.440   &     &     19.582 \\
WDFS2329+00 &   & 18.182 &  18.164 (1) &  18.145 &  18.520 (1) &  18.519 &  18.823 (2) &  18.819 &  19.075 (4)\phantom{1} &  19.087 &  19.247 (15) &  19.217 \\
WDFS2351+37 &     & 17.822 &     &     18.074  &     &     18.502   &     &     18.822   &     &     19.100   &     &     19.241 \\
\enddata 
\end{deluxetable*}

 DECaLS observed magnitudes are from DECaLS DR9 \citep{DECALSDR9}. Uncertainties are estimated as $2.5 / \lg{10} / \texttt{snr}$. Results are in Table~\ref{tab:LS_Residuals}, and in Figure~\ref{fig:LS_Resid}.

\begin{deluxetable*}{ccccccc}
\tablecaption{Observed and Synthetic Photometry in $grz$ DECaLS Survey Bands. Units are AB mag.\label{tab:LS_Residuals}}
\scriptsize
\tablehead{
\colhead{Object}   &  \colhead{$g$}   &  \colhead{$g$} &   \colhead{$r$}
&  \colhead{$r$}     &  \colhead{$z$}
&  \colhead{$z$}   \\
\colhead{}   &  \colhead{Obs.}   &  \colhead{Synth.} &   \colhead{Obs.}
&  \colhead{Synth.}   &  \colhead{Obs.}   &  \colhead{Synth.}  }
\startdata
G191B2B   &    &    11.502      &    & 12.092    &    &    12.820   \\
GD153   &    &    13.096      &    & 13.655     &    &    14.368    \\
GD71   &    &    12.797      &    & 13.335    &    &    14.035   \\
WDFS0103-00 & 19.091 (2) & 19.088 & 19.606 (3) & 19.624 & 20.291 (9)\phantom{1} & 20.309 \\
WDFS0122-30 & 18.464 (1) & 18.458 & 18.935 (1) & 18.982 & 19.666 (5)\phantom{1} & 19.669 \\
WDFS0228-08 & 19.784 (3) & 19.827 & 20.182 (4) & 20.221 & 20.809 (13) & 20.802 \\
WDFS0238-36 & 18.077 (1) & 18.095 & 18.438 (1) & 18.485 & 19.035 (3)\phantom{1} & 19.063 \\
WDFS0248+33 &  & 18.370 &  & 18.792 &  & 19.385 \\
WDFS0458-56 & 17.767 (1) & 17.754 & 18.227 (1) & 18.273 & 18.931 (3)\phantom{1} & 18.959 \\
WDFS0541-19 & 18.259 (1) & 18.272 & 18.648 (1) & 18.677 & 19.245 (4)\phantom{1} & 19.271 \\
WDFS0639-57 & 18.156 (2) & 18.177 & 18.632 (3) & 18.692 & 19.302 (6)\phantom{1} & 19.360 \\
WDFS0727+32 & 18.012 (2) & 17.999 & 18.491 (3) & 18.511 & 19.160 (5)\phantom{1} & 19.176 \\
WDFS0815+07 & 19.734 (4) & 19.721 & 20.225 (8) & 20.242 & 20.896 (19) & 20.921 \\
WDFS0956-38 &  & 17.851 &  & 18.228 &  & 18.799 \\
WDFS1024-00 & 18.909 (2) & 18.911 & 19.322 (4) & 19.365 & 19.981 (9)\phantom{1} & 19.989 \\
WDFS1055-36 &  & 18.008 &  & 18.483 &  & 19.133 \\
WDFS1110-17 &  & 17.861 &  & 18.368 &  & 19.032 \\
WDFS1111+39 & 18.407 (3) & 18.420 & 18.921 (4) & 18.995 & 19.696 (9)\phantom{1} & 19.715 \\
WDFS1206+02 & 18.664 (2) & 18.673 & 19.075 (3) & 19.114 & 19.706 (6)\phantom{1} & 19.737 \\
WDFS1206-27 &  & 16.476 &  & 16.973 &  & 17.638 \\
WDFS1214+45 & 17.743 (2) & 17.759 & 18.231 (3) & 18.294 & 18.950 (5)\phantom{1} & 18.990 \\
WDFS1302+10 & 17.026 (1) & 17.039 & 17.533 (2) & 17.570 & 18.229 (4)\phantom{1} & 18.258 \\
WDFS1314-03 & 19.102 (2) & 19.100 & 19.597 (4) & 19.622 & 20.262 (12) & 20.302 \\
WDFS1434-28 &  & 17.973 &  & 18.328 &  & 18.875 \\
WDFS1514+00 & 15.683 (.4) & 15.707 & 16.134 (.4) & 16.169 & 16.778 (1)\phantom{1} & 16.808 \\
WDFS1535-77 &  & 16.553 &  & 17.112 &  & 17.823 \\
WDFS1557+55 & 17.433 (1) & 17.470 & 17.975 (3) & 18.044 & 18.728 (4)\phantom{1} & 18.761 \\
WDFS1638+00 & 18.848 (2) & 18.837 & 19.315 (5) & 19.336 & 19.971 (12) & 19.988 \\
WDFS1814+78 & 16.570 (1) & 16.542 & 17.012 (1) & 17.063 & 17.743 (2)\phantom{1} & 17.750 \\
WDFS1837-70 &  & 17.771 &  & 18.142 &  & 18.708 \\
WDFS1930-52 &  & 17.482 &  & 17.981 &  & 18.643 \\
WDFS2101-05 & 18.638 (1) & 18.655 & 19.083 (5) & 19.114 & 19.729 (7)\phantom{1} & 19.748 \\
WDFS2317-29 &  & 18.345 &  & 18.802 &  & 19.440 \\
WDFS2329+00 & 18.133 (1) & 18.145 & 18.487 (1) & 18.519 & 19.061 (3)\phantom{1} & 19.087 \\
WDFS2351+37 &  & 18.074 &  & 18.502 &  & 19.100 \\
\enddata 
\end{deluxetable*}

PaNSTARRS1 observed PSF magnitudes are from the mean table
of the DR2 \citep{Flewellingetal2020}. Results are in Table~\ref{tab:PS1_Residuals}, and in Figure~\ref{fig:PSResid}.

\begin{deluxetable*}{ccccccccc}
\tablecaption{Observed and Synthetic Photometry in Pan-STARRS1 $griz$ Bands. Units are AB mag.\label{tab:PS1_Residuals}}
\scriptsize
\tablehead{
\colhead{Object}   &  \colhead{$g$}   &  \colhead{$g$} &   \colhead{$r$}
&  \colhead{$r$} &  \colhead{$i$}   &  \colhead{$i$}    &  \colhead{$z$}
&  \colhead{$z$}   \\
\colhead{}   &  \colhead{Obs.}   &  \colhead{Synth.} &   \colhead{Obs.}
&  \colhead{Synth.}   &  \colhead{Obs.}   &  \colhead{Synth.}   &  \colhead{Obs.}   &  \colhead{Synth.}}
\startdata
G191B2B  &  & 11.513 &  & 11.987 &  & 12.352 &  & 12.569\\
GD153  &  & 13.107 &  & 13.559 &  & 13.917 &  & 14.141\\
GD71  &  & 12.808 &  & 13.244 &  & 13.596 &  & 13.821\\
WDFS0103-00  & 19.093 (9)\phantom{1} & 19.100 & 19.570 (18) & 19.533 & 19.979 (18)\phantom{1} & 19.875 & 20.130 (69)\phantom{1} & 20.089\\
WDFS0122-30  &  & 18.470 &  & 18.895 &  & 19.239 &  & 19.462\\
WDFS0228-08  & 19.837 (13) & 19.837 & 20.188 (56) & 20.159 & 20.523 (34)\phantom{1} & 20.447 & 20.803 (88)\phantom{1} & 20.656\\
WDFS0238-36  &  & 18.106 &  & 18.423 &  & 18.709 &  & 18.917\\
WDFS0248+33  & 18.351 (7)\phantom{1} & 18.382 & 18.699 (8)\phantom{1} & 18.726 & 18.972 (12)\phantom{1} & 19.020 & 19.198 (33)\phantom{1} & 19.223\\
WDFS0458-56  &  & 17.766 &  & 18.187 &  & 18.531 &  & 18.757\\
WDFS0541-19  &  & 18.282 &  & 18.614 &  & 18.911 &  & 19.126\\
WDFS0639-57  &  & 18.189 &  & 18.605 &  & 18.938 &  & 19.150\\
WDFS0727+32  & 18.018 (11) & 18.011 & 18.475 (12) & 18.425 & 18.806 (11)\phantom{1} & 18.757 & 19.127 (30)\phantom{1} & 18.969\\
WDFS0815+07  & 19.781 (38) & 19.733 & 20.328 (42) & 20.156 & 20.625 (67)\phantom{1} & 20.497 & 20.710 (165) & 20.718\\
WDFS0956-38  &  & 17.861 &  & 18.169 &  & 18.453 &  & 18.663\\
WDFS1024-00  & 18.885 (10) & 18.923 & 19.292 (26) & 19.292 & 19.440 (102) & 19.601 & 19.758 (26)\phantom{1} & 19.810\\
WDFS1055-36  &  & 18.019 &  & 18.405 &  & 18.730 &  & 18.949\\
WDFS1110-17  & 17.895 (4)\phantom{1} & 17.874 & 18.302 (10) & 18.283 & 18.607 (13)\phantom{1} & 18.614 & 18.957 (20)\phantom{1} & 18.828\\
WDFS1111+39  & 18.412 (14) & 18.431 & 18.886 (8)\phantom{1} & 18.895 & 19.260 (11)\phantom{1} & 19.254 & 19.586 (25)\phantom{1} & 19.473\\
WDFS1206+02  & 18.693 (12) & 18.684 & 19.096 (26) & 19.044 & 19.388 (19)\phantom{1} & 19.356 & 19.645 (31)\phantom{1} & 19.574\\
WDFS1206-27  &  & 16.488 &  & 16.891 &  & 17.223 &  & 17.441\\
WDFS1214+45  & 17.779 (6)\phantom{1} & 17.770 & 18.236 (7)\phantom{1} & 18.203 & 18.569 (10)\phantom{1} & 18.553 & 18.849 (18)\phantom{1} & 18.777\\
WDFS1302+10  & 17.052 (4)\phantom{1} & 17.051 & 17.494 (5)\phantom{1} & 17.480 & 17.858 (6)\phantom{1}\phantom{1} & 17.824 & 18.114 (9)\phantom{1}\phantom{1} & 18.043\\
WDFS1314-03  & 19.078 (13) & 19.112 & 19.556 (23) & 19.535 & 19.887 (29)\phantom{1} & 19.874 & 20.240 (60)\phantom{1} & 20.091\\
WDFS1434-28  &  & 17.983 &  & 18.272 &  & 18.542 &  & 18.744\\
WDFS1514+00  & 15.720 (3)\phantom{1} & 15.718 & 16.101 (4)\phantom{1} & 16.094 & 16.434 (2)\phantom{1}\phantom{1} & 16.412 & 16.715 (5)\phantom{1}\phantom{1} & 16.630\\
WDFS1535-77  &  & 16.565 &  & 17.013 &  & 17.367 &  & 17.587\\
WDFS1557+55  & 17.487 (5)\phantom{1} & 17.482 & 17.958 (6)\phantom{1} & 17.944 & 18.356 (5)\phantom{1}\phantom{1} & 18.303 & 18.647 (13)\phantom{1} & 18.520\\
WDFS1638+00  & 18.860 (15) & 18.849 & 19.314 (23) & 19.252 & 19.611 (14)\phantom{1} & 19.577 & 19.816 (48)\phantom{1} & 19.786\\
WDFS1814+78  & 16.573 (5)\phantom{1} & 16.553 & 17.007 (3)\phantom{1} & 16.976 & 17.358 (4)\phantom{1}\phantom{1} & 17.321 & 17.651 (9)\phantom{1}\phantom{1} & 17.546\\
WDFS1837-70  &  & 17.781 &  & 18.085 &  & 18.365 &  & 18.574\\
WDFS1930-52  &  & 17.495 &  & 17.899 &  & 18.230 &  & 18.446\\
WDFS2101-05  & 18.652 (8)\phantom{1} & 18.667 & 19.052 (8)\phantom{1} & 19.040 & 19.410 (20)\phantom{1} & 19.356 & 19.703 (38)\phantom{1} & 19.572\\
WDFS2317-29  &  & 18.356 &  & 18.729 &  & 19.050 &  & 19.273\\
WDFS2329+00  & 18.134 (5)\phantom{1} & 18.154 & 18.452 (5)\phantom{1} & 18.460 & 18.772 (10)\phantom{1} & 18.741 & 19.003 (13)\phantom{1} & 18.948\\
WDFS2351+37  & 18.085 (3)\phantom{1} & 18.086 & 18.447 (11) & 18.434 & 18.776 (12)\phantom{1} & 18.729 & 19.100 (38)\phantom{1} & 18.930\\
\enddata

\end{deluxetable*}

The observed SDSS data for the CALSPEC DAWD standards are from \citet{Holberg2006}, modified to correct a typographical error in the $i$-band flux of GD153 (Holberg, private communication). The observed magnitudes for the fainter DAWDs come from SDSS DR7 (\citet{SDSSDR7}).

\begin{deluxetable*}{ccccccccccc}
\tablecaption{Observed and Synthetic Photometry in SDSS $ugriz$ Bands. Units are AB mag.\label{tab:SDSS_Residuals}}
\scriptsize
\tablehead{
\colhead{Object}  &  \colhead{$u$}   &  \colhead{$u$}  &  \colhead{$g$}   &  \colhead{$g$} &   \colhead{$r$}
&  \colhead{$r$}   &  \colhead{$i$}   &  \colhead{$i$}   &  \colhead{$z$}
&  \colhead{$z$}   \\
\colhead{}   &  \colhead{Obs.}   &  \colhead{Synth.} &   \colhead{Obs.}
&  \colhead{Synth.}   &  \colhead{Obs.}   &  \colhead{Synth.}   &  \colhead{Obs.}
&  \colhead{Synth.}   &  \colhead{Obs.}   &  \colhead{Synth.}}
\startdata
G191B2B & 11.033 (16) & 10.997 & 11.470 (4)\phantom{1} & 11.479 & 12.007 (7)\phantom{1} & 12.020 & 12.388 (4)\phantom{1} & 12.407 & 12.740 (6)\phantom{1}\phantom{1} & 12.766\\
GD153 & 12.700 (40) & 12.667 & 13.022 (12) & 13.075 & 13.573 (11) & 13.585 & 13.950 (9)\phantom{1} & 13.961 & 14.307 (16)\phantom{1} & 14.315\\
GD71 & 12.438 (17) & 12.430 & 12.752 (1)\phantom{1} & 12.778 & 13.241 (12) & 13.266 & 13.611 (4)\phantom{1} & 13.633 & 13.973 (18)\phantom{1} & 13.984\\
WDFS0103-00 & 18.643 (22) & 18.633 & 19.060 (11) & 19.067 & 19.509 (17) & 19.558 & 19.906 (32) & 19.918 & 20.198 (158) & 20.258\\
WDFS0122-30 &  & 18.091 &  & 18.440 &  & 18.916 &  & 19.275 &  & 19.619\\
WDFS0228-08 & 19.798 (41) & 19.765 & 19.769 (15) & 19.820 & 20.150 (25) & 20.166 & 20.367 (42) & 20.461 & 21.197 (410) & 20.760\\
WDFS0238-36 &  & 18.033 &  & 18.089 &  & 18.430 &  & 18.723 &  & 19.020\\
WDFS0248+33 & 18.105 (14) & 18.118 & 18.330 (7)\phantom{1} & 18.356 & 18.690 (9)\phantom{1} & 18.736 & 18.921 (14) & 19.040 & 19.213 (53)\phantom{1} & 19.341\\
WDFS0458-56 &  & 17.441 &  & 17.738 &  & 18.206 &  & 18.564 &  & 18.909\\
WDFS0541-19 &  & 18.271 &  & 18.265 &  & 18.620 &  & 18.923 &  & 19.228\\
WDFS0639-57 &  & 17.742 &  & 18.157 &  & 18.627 &  & 18.977 &  & 19.310\\
WDFS0727+32 & 17.564 (11) & 17.570 & 17.962 (6)\phantom{1} & 17.979 & 18.455 (8)\phantom{1} & 18.447 & 18.780 (13) & 18.795 & 19.042 (57)\phantom{1} & 19.126\\
WDFS0815+07 & 19.385 (28) & 19.358 & 19.651 (12) & 19.701 & 20.177 (23) & 20.176 & 20.528 (37) & 20.532 & 20.540 (153) & 20.872\\
WDFS0956-38 &  & 17.910 &  & 17.847 &  & 18.174 &  & 18.462 &  & 18.758\\
WDFS1024-00 & 18.586 (17) & 18.592 & 18.839 (9)\phantom{1} & 18.896 & 19.292 (13) & 19.306 & 19.592 (21) & 19.627 & 19.759 (79)\phantom{1} & 19.942\\
WDFS1055-36 &  & 17.739 &  & 17.994 &  & 18.421 &  & 18.756 &  & 19.085\\
WDFS1110-17 & 17.480 (11) & 17.448 & 17.825 (6)\phantom{1} & 17.843 & 18.294 (8)\phantom{1} & 18.304 & 18.612 (12) & 18.650 & 18.909 (43)\phantom{1} & 18.983\\
WDFS1111+39 & 17.960 (13) & 17.933 & 18.374 (7)\phantom{1} & 18.398 & 18.905 (10) & 18.925 & 19.264 (17) & 19.305 & 19.628 (68)\phantom{1} & 19.661\\
WDFS1206+02 &  & 18.553 &  & 18.663 &  & 19.054 &  & 19.374 &  & 19.692\\
WDFS1206-27 &  & 16.130 &  & 16.459 &  & 16.909 &  & 17.254 &  & 17.588\\
WDFS1214+45 & 17.358 (9)\phantom{1} & 17.378 & 17.700 (5)\phantom{1} & 17.740 & 18.197 (7)\phantom{1} & 18.226 & 18.540 (12) & 18.591 & 18.763 (34)\phantom{1} & 18.939\\
WDFS1302+10 & 16.637 (8)\phantom{1} & 16.619 & 16.982 (4)\phantom{1} & 17.019 & 17.468 (6)\phantom{1} & 17.503 & 17.842 (7)\phantom{1} & 17.864 & 18.146 (28)\phantom{1} & 18.207\\
WDFS1314-03 &  & 18.684 &  & 19.081 &  & 19.557 &  & 19.912 &  & 20.251\\
WDFS1434-28 &  & 18.021 &  & 17.969 &  & 18.276 &  & 18.551 &  & 18.836\\
WDFS1514+00 & 15.475 (4)\phantom{1} & 15.467 & 15.663 (3)\phantom{1} & 15.694 & 16.089 (4)\phantom{1} & 16.108 & 16.412 (4)\phantom{1} & 16.437 & 16.728 (12)\phantom{1} & 16.761\\
WDFS1535-77 &  & 16.073 &  & 16.533 &  & 17.042 &  & 17.415 &  & 17.770\\
WDFS1557+55 & 16.982 (8)\phantom{1} & 16.985 & 17.438 (5)\phantom{1} & 17.448 & 17.985 (7)\phantom{1} & 17.974 & 18.344 (10) & 18.353 & 18.685 (38)\phantom{1} & 18.708\\
WDFS1638+00 &  & 18.412 &  & 18.817 &  & 19.273 &  & 19.613 &  & 19.939\\
WDFS1814+78 &  & 16.213 &  & 16.525 &  & 16.996 &  & 17.355 &  & 17.700\\
WDFS1837-70 &  & 17.838 &  & 17.768 &  & 18.088 &  & 18.374 &  & 18.667\\
WDFS1930-52 &  & 17.121 &  & 17.464 &  & 17.917 &  & 18.262 &  & 18.594\\
WDFS2101-05 & 18.460 (17) & 18.414 & 18.651 (9)\phantom{1} & 18.642 & 19.046 (12) & 19.053 & 19.388 (22) & 19.380 & 19.791 (93)\phantom{1} & 19.701\\
WDFS2317-29 &  & 18.223 &  & 18.334 &  & 18.740 &  & 19.069 &  & 19.394\\
WDFS2329+00 &  & 18.161 &  & 18.140 &  & 18.465 &  & 18.751 &  & 19.045\\
WDFS2351+37 & 17.771 (11) & 17.749 & 18.022 (6)\phantom{1} & 18.059 & 18.437 (8)\phantom{1} & 18.446 & 18.757 (11) & 18.752 & 19.007 (46)\phantom{1} & 19.055\\
\enddata

\end{deluxetable*}

Gaia observed magnitudes are from DR3 \citep{GAIADR3}. For comparison to DR3 magnitudes, our synthetic magnitudes are transformed from AB to Vega using passband data from \citet{SVO20}. Results are in Table \ref{tab:Gaia_Residuals}, and Figure \ref{fig:GaiaResid}.

\begin{deluxetable*}{ccccccc}
\tablecaption{Observed and Synthetic Photometry in Gaia Bands. Units are Vega mag.\label{tab:Gaia_Residuals}}
\scriptsize
\tablehead{
\colhead{Object}   &  \colhead{$G$}   &  \colhead{$G$} &   \colhead{$RP$}
&  \colhead{$RP$}     &  \colhead{$BP$}
&  \colhead{$BP$}   \\
\colhead{}   &  \colhead{Obs.}   &  \colhead{Synth.} &   \colhead{Obs.}
&  \colhead{Synth.}   &  \colhead{Obs.}   &  \colhead{Synth.}  }
\startdata
G191B2B & 11.718 (3) & 11.715 & 12.071 (4)\phantom{1}\phantom{1} & 12.054 & 11.546 (3)\phantom{1} & 11.539\\
GD153 & 13.311 (3) & 13.300 & 13.632 (4)\phantom{1}\phantom{1} & 13.611 & 13.151 (3)\phantom{1} & 13.139\\
GD71 & 13.000 (3) & 12.996 & 13.305 (4)\phantom{1}\phantom{1} & 13.286 &  12.853 (3)\phantom{1} & 12.845\\
WDFS0103-00 & 19.302 (3) & 19.279 & 19.672 (53)\phantom{1} & 19.566 & 19.164 (33) & 19.123\\
WDFS0122-30 & 18.664 (1) & 18.650 & 19.010 (32)\phantom{1} & 18.927 & 18.532 (14) & 18.504\\
WDFS0228-08 & 19.975 (6) & 19.969 & 20.068 (171) & 20.120 & 19.820 (75) & 19.886\\
WDFS0238-36 & 18.236 (1) & 18.235 & 18.386 (25)\phantom{1} & 18.381 & 18.188 (14) & 18.154\\
WDFS0248+33 & 18.521 (2) & 18.516 & 18.742 (43)\phantom{1} & 18.691 & 18.423 (21) & 18.411\\
WDFS0458-56 & 17.959 (1) & 17.948 & 18.251 (37)\phantom{1} & 18.219 & 17.847 (12) & 17.807\\
WDFS0541-19 & 18.433 (2) & 18.423 & 18.607 (26)\phantom{1} & 18.583 & 18.349 (14) & 18.340\\
WDFS0639-57 & 18.375 (2) & 18.359 & 18.702 (41)\phantom{1} & 18.625 & 18.269 (15) & 18.211\\
WDFS0727+32 & 18.189 (2) & 18.180 & 18.452 (40)\phantom{1} & 18.443 & 18.043 (13) & 18.033\\
WDFS0815+07 & 19.932 (5) & 19.911 & 20.248 (129) & 20.183 & 19.787 (51) & 19.766\\
WDFS0956-38 & 18.002 (1) & 17.990 & 18.157 (15)\phantom{1} & 18.124 & 17.945 (7)\phantom{1} & 17.919\\
WDFS1024-00 & 19.083 (3) & 19.070 & 19.234 (53)\phantom{1} & 19.279 & 18.996 (33) & 18.950\\
WDFS1055-36 & 18.196 (1) & 18.182 & 18.453 (18)\phantom{1} & 18.412 & 18.121 (11) & 18.058\\
WDFS1110-17 & 18.048 (1) & 18.041 & 18.372 (30)\phantom{1} & 18.300 & 17.911 (9)\phantom{1} & 17.897\\
WDFS1111+39 & 18.644 (2) & 18.628 & 19.067 (53)\phantom{1} & 18.953 & 18.485 (20) & 18.457\\
WDFS1206+02 & 18.850 (2) & 18.838 & 19.066 (43)\phantom{1} & 19.032 & 18.746 (33) & 18.735\\
WDFS1206-27 & 16.667 (1) & 16.656 & 16.930 (10)\phantom{1} & 16.907 & 16.543 (3)\phantom{1} & 16.519\\
WDFS1214+45 & 17.979 (1) & 17.955 & 18.226 (26)\phantom{1} & 18.243 & 17.836 (8)\phantom{1} & 17.804\\
WDFS1302+10 & 17.239 (1) & 17.230 & 17.542 (13)\phantom{1} & 17.514 & 17.099 (4)\phantom{1} & 17.078\\
WDFS1314-03 & 19.307 (3) & 19.287 & 19.745 (83)\phantom{1} & 19.562 & 19.252 (31) & 19.138\\
WDFS1434-28 & 18.103 (2) & 18.099 & 18.352 (30)\phantom{1} & 18.211 & 18.070 (29) & 18.036\\
WDFS1514+00 & 15.884 (1) & 15.876 & 16.111 (6)\phantom{1}\phantom{1} & 16.093 & 15.775 (3)\phantom{1} & 15.758\\
WDFS1535-77 & 16.765 (1) & 16.754 & 17.095 (7)\phantom{1}\phantom{1} & 17.067 & 16.600 (3)\phantom{1} & 16.588\\
WDFS1557+55 & 17.691 (1) & 17.678 & 18.036 (25)\phantom{1} & 18.001 & 17.527 (10) & 17.507\\
WDFS1638+00 & 19.025 (2) & 19.011 & 19.362 (41)\phantom{1} & 19.261 & 18.912 (21) & 18.869\\
WDFS1814+78 & 16.745 (1) & 16.735 & 17.033 (8)\phantom{1}\phantom{1} & 17.009 & 16.612 (6)\phantom{1} & 16.593\\
WDFS1837-70 & 17.910 (1) & 17.907 & 18.081 (17)\phantom{1} & 18.035 & 17.853 (12) & 17.839\\
WDFS1930-52 & 17.673 (1) & 17.662 & 17.942 (22)\phantom{1} & 17.913 & 17.547 (7)\phantom{1} & 17.524\\
WDFS2101-05 & 18.827 (2) & 18.822 & 19.096 (38)\phantom{1} & 19.035 & 18.739 (16) & 18.706\\
WDFS2317-29 & 18.526 (2) & 18.518 & 18.809 (42)\phantom{1} & 18.728 & 18.444 (30) & 18.410\\
WDFS2329+00 & 18.292 (2) & 18.280 & 18.417 (31)\phantom{1} & 18.412 & 18.237 (21) & 18.208\\
WDFS2351+37 & 18.235 (2) & 18.219 & 18.500 (26)\phantom{1} & 18.403 & 18.122 (20) & 18.107\\
\enddata

\end{deluxetable*}

Our synthetic magnitudes are anchored on observations made above the atmosphere. 
 With the exception of Gaia, all other surveys in this comparison are from ground- based surveys, which have required accounting for the constantly changing extinction from  the terrestrial atmosphere.  We hope that the network of spectrophotometric standard stars presented in this paper will be a useful tool for resolving any discrepancies between different surveys.  Beyond this, we have not attempted to ascribe specific causes for the discrepancies between our synthetic magnitudes for these surveys and those published by the surveys themselves.  To do so would require significant expertise from the survey teams, and could form a focus for future work. Similarly, the impact of any downstream scientific results from any re-calibration of existing surveys is best left to the discretion of topical experts, and the specific problems they have at hand.

\section{Conclusions}
\label{sec:concl}
\subsection{Major results of this paper}
This and prior papers (\citetalias{Narayan19}, \citetalias{Calamida19}, and \citetalias{Calamida22}) work towards an all-sky network of faint spectrophotometric standard stars (called WDFS) composed of hot DAWDs whose fluxes are tied to the three primary CALSPEC standards.  Our initial goal of 1\% absolute and 0.5\% relative flux calibration in the visual band is realized and our network of 35 White Dwarf Flux Standards (32 WDFS and 3 CALSPEC) is available for general use. In addition, each stellar reddened SED provides predicted
magnitudes in several common survey systems (\ref{sec:surveymags}). This set of stars covers the entire sky, such that, at any time, two or more  standards are above 2 airmasses, at any ground-based observatory. 
Because the HST/WFC3 photometry that defines our system is above the
atmosphere, ground-based, atmospheric extinction problems do not exist. Our standards are suitable for many of the existing and future large telescope surveys. 

The conversion of our derived SEDs to magnitudes in ground based surveys must necessarily include the filter functions of the surveys, which are often available \citep{SVO20, SVO12}. Our published SEDs can be convolved with any filter function for any telescope. If these functions are not known, a later paper in this series, using parallel ACS images, can define color terms for conversion of native ground-based magnitudes to 
magnitudes on the space-based system.

Item 4 of the next section discusses extrapolations of our
SEDs shortward of the HST F275W passband and longward of the infrared F160W passband.

\subsection{Possible Improvements and Enhancements}
 

The following items discuss some limitations of this sample and some possible future improvements. 
\begin{enumerate}            
\item By their very nature, our white dwarf stars have blue SEDs.  If our standards are used for the calibration of broadband photometry for much redder stars, the extreme color of our WDFS stars could be problematic.  In this regard, our ACS fields (in prep) will provide photometry from the ACS/HST fields that were observed in parallel with the WFC3 observations\footnote{GOs 13711, 15113, PI: A. Saha}. These fields include approximately 100-200 stars of different spectral types within 4-6 arcminutes of our WDFS stars and
should be helpful in photometrically linking our blue standards DAWDs with redder stars.

\item Our absolute photometry is tied to CALSPEC, which has an estimated uncertainty of 1\% and is ultimately linked to the monochromatic flux of Vega at 5556 \AA\ and Sirius in the IR \citep{Bohlin2014b, Bohlin2020}, which have their own uncertainties. Ongoing and proposed ground-based and space-based efforts seek to establish stellar calibrations with respect to NIST (National Institute of Standards and Technology) laboratory radiometry with 0.5\% absolute and 0.3\% relative uncertainties in the visual.  When available, these improvements can be applied to our existing WDFS by making global corrections at the few tenths of a percent level.

\item Future expansion of our standard star network is possible.  The size of our network was ultimately dictated by the observational effort required to locate and validate suitable candidates, as well as monitoring each star for photometric stability, by obtaining the spectroscopic time on large telescopes, and by obtaining the WFC3 photometry in six bands.  Future efforts can rely on current deep multifiber spectroscopic surveys to identify large numbers of suitable candidate DAWD standards.  Likewise, multi-epoch photometry from Gaia and RST can verify photometric stability of these candidates.  However, our unique step that uses WFC3 photometry will be impossible after HST is decommissioned.

\item Extensions of wavelength range. Our fluxes are well defined over the wavelength range 2750 Å to 1.6$\mu m$ by WFC3 photometry. In order to validate our treatment of interstellar extinction and our model fluxes below 2750 Å, a new STIS/HST program obtained observations\footnote{GO 16764, PI: G. Narayan} of about two thirds of our WDFS stars in the UV down to 1150 Å. Preliminary results from this program show that our optically estimated values of \av\ predict the observed UV fluxes for most stars to a precision better than 3\%. Outliers might be explained by adjusting \av\ and $R_V$, the ratio of absolute extinction \av\ to selective extinction E(B-V), within our uncertainties.
There are no observational tests longward of the F160W passband, but there is exquisite agreement between models and observations at shorter wavelengths.

\item The placement of our WDFS SEDs on the CALSPEC absolute flux scale has several inaccuracies at the percent level that will be addressed in our next paper. First, our model SEDs are in air above 2000 Å.
The air to vacuum correction that is applied to our final SEDs is adequate, except for a small unphysical discontinuity at 2000 Å, but these
models extend to only 1350 Å in the FUV and $2.7\mu$ in the IR. CALSPEC now utilizes the NLTE grids computed by I. Hubeny and T. Rauch \citep{Bohlin2020} that cover 900 Å to 30$\mu$ with native vacuum wavelengths and show emission lines at HI line centers where these features are actually observed. Furthermore, these newer NLTE grids
include many more IR lines, including some important features like Paschen $\alpha$. Measurements of radial velocities would improve our
model SEDs slightly.

Perhaps the most important WDFS future improvement will be to place the absolute
fluxes on the updated scale of 
\citet{Bohlin2020}. Figure \ref{fig:CALSPECGD153} illustrates quantitative comparisons with these old and new flux scales \citep[GD153 is corrected for the published radial velocity of 8.3 km~s$^{-1}$][]{Napiwotzki2020}.
The blue curve
compares our GD153 model with the 2014 gd153\_mod\_010.fits and is generally within 1\% of agreement, except in the line profiles
and at the shortest wavelengths. Because our flux scale is based on the 2014 models and the WFC3 calibration of \citet{Calamida19} that is used in in our previous papers, the blue trace represents the small offset between our flux scale and the 2014 CALSPEC flux system.  The difference between the red and blue is the amount of change in 2020 to the gd153\_mod\_011.fits model of \citet{Bohlin2020}. 

\end{enumerate}

\section{Data Availability}
We have created a Zenodo url\footnote{ \url{https://doi.org/10.5281/zenodo.7713704}} that contains the SEDs derived in this paper, the WFC3 passbands we employed to create magnitudes in commonly used systems, and derived parameters for each star (Table 5 in the text), along with all of the other tables. These data and the corrected, "c",
magnitudes in Table 2 define our magnitudes in Tables 6-10. DAWD-based
magnitudes for an arbitrary system/telescope can be derived given atmospheric
transmission plus filter, mirror, and CCD efficiencies.

\acknowledgments

This paper uses tables of observed photometry from the DES survey, DECaLS, PaNSTARRS, SDSS, and Gaia. Spectroscopy was obtained at SOAR for the Southern DAWD,
and at Gemini and the MMT for the Northern and equatorial DAWD. Time series photometry was obtained at Las Cumbres Observatory. Formal acknowledgments are below.

We acknowledge support from STSCI/HST: HST-GO-12967, HST-GO-13711, HST-GO-15113. EO was also partially supported
by NSF grants  AST-1815767 and AST-1313006. 

We thank Thomas Rauch, David Buckley, and Clare Shanahan. We also thank Nicola Gentile-Fusillo and Roberto Raddi for catalogs of southern white-dwarf candidates.

EO wishes to remember Jill Bechtold here. AS dedicates his efforts towards this project to the memory of his late advisor Dr. J.B. Oke, whose measurements of the absolute flux distribution of Vega pioneered the application of spectrophotometry in astronomy. 

Spectroscopic observations reported here were obtained at the MMT Observatory, a joint facility of the University of Arizona and the Smithsonian Institution.

Spectra were also obtained at SOAR and at Gemini:
Based in part on observations obtained at the Southern Astrophysical Research (SOAR) telescope, which is a joint project of the Ministério da Ciência, Tecnologia e Inovações do Brasil (MCTI/LNA), the US National Science Foundation’s NOIRLab, the University of North Carolina at Chapel Hill (UNC), and Michigan State University (MSU).

Based on observations obtained at the international Gemini Observatory, a program of NSF’s NOIRLab, which is managed by the Association of Universities for Research in Astronomy (AURA) under a cooperative agreement with the National Science Foundation on behalf of the Gemini Observatory partnership: the National Science Foundation (United States), National Research Council (Canada), Agencia Nacional de Investigaci\'{o}n y Desarrollo (Chile), Ministerio de Ciencia, Tecnolog\'{i}a e Innovaci\'{o}n (Argentina), Minist\'{e}rio da Ci\^{e}ncia, Tecnologia, Inova\c{c}\~{o}es e Comunica\c{c}\~{o}es (Brazil), and Korea Astronomy and Space Science Institute (Republic of Korea).

This work makes use of observations from the Las Cumbres Observatory global telescope network.

This work has made use of data from the European Space Agency (ESA) mission Gaia (https://www.cosmos.esa.int/gaia), processed by the Gaia Data Processing and Analysis Consortium (DPAC; https://www.cosmos.esa.int/web/gaia/dpac/consortium). Funding for the DPAC has been provided by national institutions, in particular the institutions participating in the Gaia Multilateral Agreement. This publication makes use of VOSA, developed under the Spanish Virtual Observatory project supported by the Spanish MINECO through grant AyA2017-84089. VOSA has been partially updated by using funding from the European Union's Horizon 2020 Research and Innovation Programme, under grant Agreement No. 776403 (EXOPLANETS-A).

The Pan-STARRS1 Surveys (PS1) and the PS1 public science archive have been made possible through contributions by the Institute for Astronomy, the University of Hawaii, the Pan-STARRS Project Office, the Max-Planck Society and its participating institutes, the Max Planck Institute for Astronomy, Heidelberg and the Max Planck Institute for Extraterrestrial Physics, Garching, The Johns Hopkins University, Durham University, the University of Edinburgh, the Queen's University Belfast, the Harvard-Smithsonian Center for Astrophysics, the Las Cumbres Observatory Global Telescope Network Incorporated, the National Central University of Taiwan, the Space Telescope Science Institute, the National Aeronautics and Space Administration under Grant No. NNX08AR22G issued through the Planetary Science Division of the NASA Science Mission Directorate, the National Science Foundation Grant No. AST-1238877, the University of Maryland, Eotvos Lorand University (ELTE), the Los Alamos National Laboratory, and the Gordon and Betty Moore Foundation.

This project used public archival data from the Dark Energy Survey (DES). Funding for the DES Projects has been provided by the U.S. Department of Energy, the U.S. National Science Foundation, the Ministry of Science and Education of Spain, the Science and Technology FacilitiesCouncil of the United Kingdom, the Higher Education Funding Council for England, the National Center for Supercomputing Applications at the University of Illinois at Urbana-Champaign, the Kavli Institute of Cosmological Physics at the University of Chicago, the Center for Cosmology and Astro-Particle Physics at the Ohio State University, the Mitchell Institute for Fundamental Physics and Astronomy at Texas A\&M University, Financiadora de Estudos e Projetos, Funda{\c c}{\~a}o Carlos Chagas Filho de Amparo {\`a} Pesquisa do Estado do Rio de Janeiro, Conselho Nacional de Desenvolvimento Cient{\'i}fico e Tecnol{\'o}gico and the Minist{\'e}rio da Ci{\^e}ncia, Tecnologia e Inova{\c c}{\~a}o, the Deutsche Forschungsgemeinschaft, and the Collaborating Institutions in the Dark Energy Survey.
The Collaborating Institutions are Argonne National Laboratory, the University of California at Santa Cruz, the University of Cambridge, Centro de Investigaciones Energ{\'e}ticas, Medioambientales y Tecnol{\'o}gicas-Madrid, the University of Chicago, University College London, the DES-Brazil Consortium, the University of Edinburgh, the Eidgen{\"o}ssische Technische Hochschule (ETH) Z{\"u}rich,  Fermi National Accelerator Laboratory, the University of Illinois at Urbana-Champaign, the Institut de Ci{\`e}ncies de l'Espai (IEEC/CSIC), the Institut de F{\'i}sica d'Altes Energies, Lawrence Berkeley National Laboratory, the Ludwig-Maximilians Universit{\"a}t M{\"u}nchen and the associated Excellence Cluster Universe, the University of Michigan, the National Optical Astronomy Observatory, the University of Nottingham, The Ohio State University, the OzDES Membership Consortium, the University of Pennsylvania, the University of Portsmouth, SLAC National Accelerator Laboratory, Stanford University, the University of Sussex, and Texas A\&M University.
Based in part on observations at Cerro Tololo Inter-American Observatory, National Optical Astronomy Observatory, which is operated by the Association of Universities for Research in Astronomy (AURA) under a cooperative agreement with the National Science Foundation.

The Legacy Surveys consist of three individual and complementary projects: the Dark Energy Camera Legacy Survey (DECaLS; Proposal ID \#2014B-0404; PIs: David Schlegel and Arjun Dey), the Beijing-Arizona Sky Survey (BASS; NOAO Prop. ID \#2015A-0801; PIs: Zhou Xu and Xiaohui Fan), and the Mayall z-band Legacy Survey (MzLS; Prop. ID \#2016A-0453; PI: Arjun Dey). DECaLS, BASS and MzLS together include data obtained, respectively, at the Blanco telescope, Cerro Tololo Inter-American Observatory, NSF’s NOIRLab; the Bok telescope, Steward Observatory, University of Arizona; and the Mayall telescope, Kitt Peak National Observatory, NOIRLab. Pipeline processing and analyses of the data were supported by NOIRLab and the Lawrence Berkeley National Laboratory (LBNL). The Legacy Surveys project is honored to be permitted to conduct astronomical research on Iolkam Du’ag (Kitt Peak), a mountain with particular significance to the Tohono O’odham Nation.

NOIRLab is operated by the Association of Universities for Research in Astronomy (AURA) under a cooperative agreement with the National Science Foundation. LBNL is managed by the Regents of the University of California under contract to the U.S. Department of Energy.

This project used data obtained with the Dark Energy Camera (DECam), which was constructed by the Dark Energy Survey (DES) collaboration. Funding for the DES Projects has been provided by the U.S. Department of Energy, the U.S. National Science Foundation, the Ministry of Science and Education of Spain, the Science and Technology Facilities Council of the United Kingdom, the Higher Education Funding Council for England, the National Center for Supercomputing Applications at the University of Illinois at Urbana-Champaign, the Kavli Institute of Cosmological Physics at the University of Chicago, Center for Cosmology and Astro-Particle Physics at the Ohio State University, the Mitchell Institute for Fundamental Physics and Astronomy at Texas A\&M University, Financiadora de Estudos e Projetos, Fundacao Carlos Chagas Filho de Amparo, Financiadora de Estudos e Projetos, Fundacao Carlos Chagas Filho de Amparo a Pesquisa do Estado do Rio de Janeiro, Conselho Nacional de Desenvolvimento Cientifico e Tecnologico and the Ministerio da Ciencia, Tecnologia e Inovacao, the Deutsche Forschungsgemeinschaft and the Collaborating Institutions in the Dark Energy Survey. The Collaborating Institutions are Argonne National Laboratory, the University of California at Santa Cruz, the University of Cambridge, Centro de Investigaciones Energeticas, Medioambientales y Tecnologicas-Madrid, the University of Chicago, University College London, the DES-Brazil Consortium, the University of Edinburgh, the Eidgenossische Technische Hochschule (ETH) Zurich, Fermi National Accelerator Laboratory, the University of Illinois at Urbana-Champaign, the Institut de Ciencies de l’Espai (IEEC/CSIC), the Institut de Fisica d’Altes Energies, Lawrence Berkeley National Laboratory, the Ludwig Maximilians Universitat Munchen and the associated Excellence Cluster Universe, the University of Michigan, NSF’s NOIRLab, the University of Nottingham, the Ohio State University, the University of Pennsylvania, the University of Portsmouth, SLAC National Accelerator Laboratory, Stanford University, the University of Sussex, and Texas A\&M University.

BASS is a key project of the Telescope Access Program (TAP), which has been funded by the National Astronomical Observatories of China, the Chinese Academy of Sciences (the Strategic Priority Research Program “The Emergence of Cosmological Structures” Grant \# XDB09000000), and the Special Fund for Astronomy from the Ministry of Finance. The BASS is also supported by the External Cooperation Program of Chinese Academy of Sciences (Grant \# 114A11KYSB20160057), and Chinese National Natural Science Foundation (Grant \# 12120101003, \# 11433005).

The Legacy Survey team makes use of data products from the Near-Earth Object Wide-field Infrared Survey Explorer (NEOWISE), which is a project of the Jet Propulsion Laboratory/California Institute of Technology. NEOWISE is funded by the National Aeronautics and Space Administration.

The Legacy Surveys imaging of the DESI footprint is supported by the Director, Office of Science, Office of High Energy Physics of the U.S. Department of Energy under Contract No. DE-AC02-05CH1123, by the National Energy Research Scientific Computing Center, a DOE Office of Science User Facility under the same contract; and by the U.S. National Science Foundation, Division of Astronomical Sciences under Contract No. AST-0950945 to NOAO.

Funding for the Sloan Digital Sky
Survey IV has been provided by the
Alfred P. Sloan Foundation, the U.S.
Department of Energy Office of
Science, and the Participating
Institutions.

SDSS-IV acknowledges support and
resources from the Center for High
Performance Computing  at the
University of Utah. The SDSS
website is www.sdss4.org.

SDSS-IV is managed by the
Astrophysical Research Consortium
for the Participating Institutions
of the SDSS Collaboration including
the Brazilian Participation Group,
the Carnegie Institution for Science,
Carnegie Mellon University, Center for
Astrophysics | Harvard \&
Smithsonian, the Chilean Participation
Group, the French Participation Group,
Instituto de Astrof\'isica de
Canarias, The Johns Hopkins
University, Kavli Institute for the
Physics and Mathematics of the
Universe (IPMU) / University of
Tokyo, the Korean Participation Group,
Lawrence Berkeley National Laboratory,
Leibniz Institut f\"ur Astrophysik
Potsdam (AIP),  Max-Planck-Institut
f\"ur Astronomie (MPIA Heidelberg),
Max-Planck-Institut f\"ur
Astrophysik (MPA Garching),
Max-Planck-Institut f\"ur
Extraterrestrische Physik (MPE),
National Astronomical Observatories of
China, New Mexico State University,
New York University, University of
Notre Dame, Observat\'ario
Nacional / MCTI, The Ohio State
University, Pennsylvania State
University, Shanghai
Astronomical Observatory, United
Kingdom Participation Group,
Universidad Nacional Aut\'onoma
de M\'exico, University of Arizona,
University of Colorado Boulder,
University of Oxford, University of
Portsmouth, University of Utah,
University of Virginia, University
of Washington, University of
Wisconsin, Vanderbilt University,
and Yale University.

We thank an anonymous referee for comments which allowed us to improve the paper.

\facilities{\emph{HST} (WFC3), \emph{SOAR}, \emph{MMT}, \emph {Gemini}, \emph{Gaia},\emph{PaNSTARRS}, \emph{SDSS}, \emph{DES},\emph{DECaLS},
 \emph{Las Cumbres Observatory}.
}

\bibliographystyle{aasjournal}
\bibliography{references}

\begin{figure*}
\begin{center}

\includegraphics[width=1.0\textwidth]{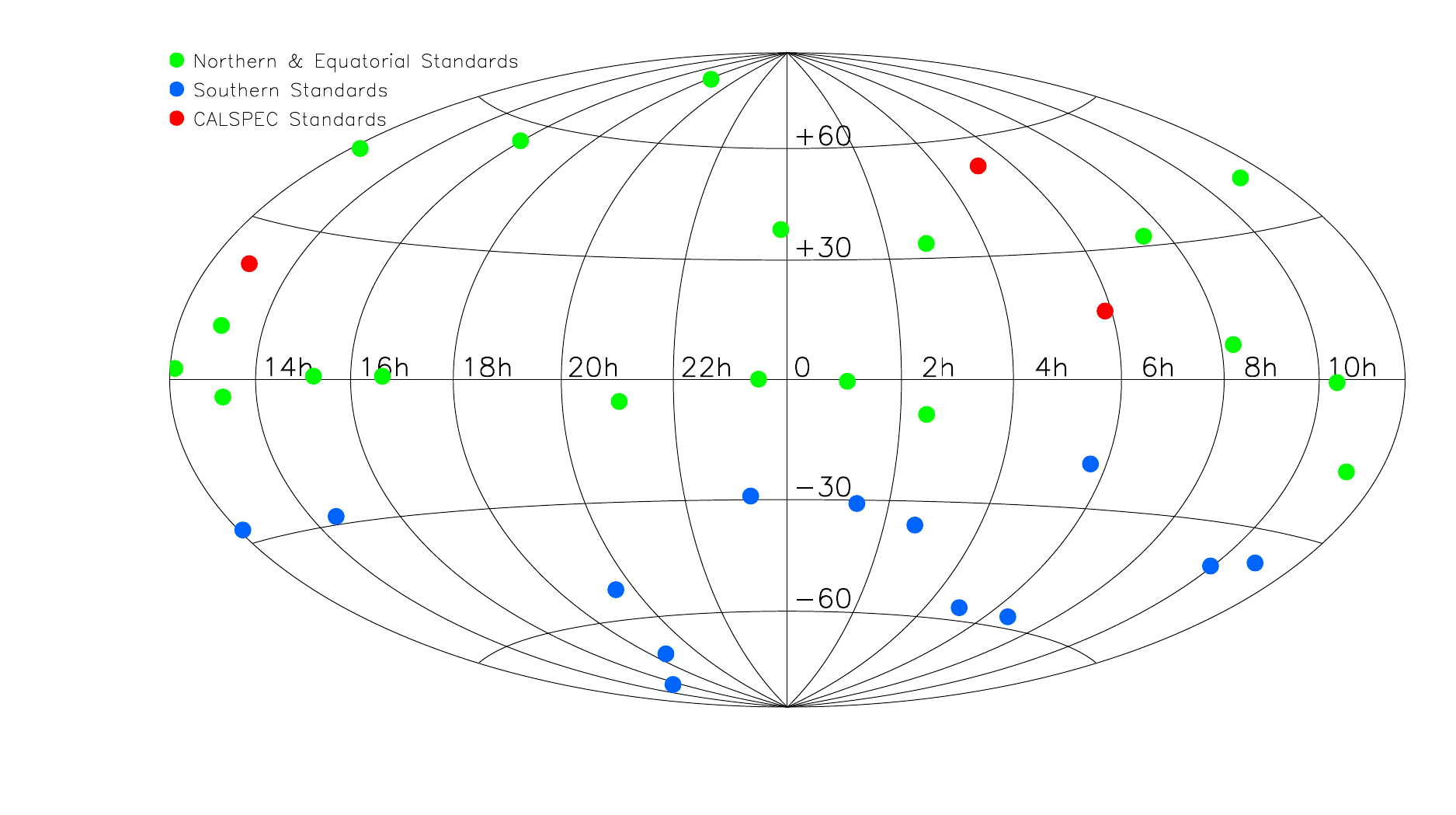}

\caption{Aitoff projection of our network of  spectrophotometric standard DAWDs illustrating the full-sky nature of our system ({\it HST} Cycles 20 and 22 in green, Cycle 25 in blue, and the three bright CALSPEC DAWDs in red). \label{fig:allsky}}
\end{center}
\end{figure*}

\begin{figure*}[h!tpb]
    \centering
    \includegraphics[width=0.85\textwidth]{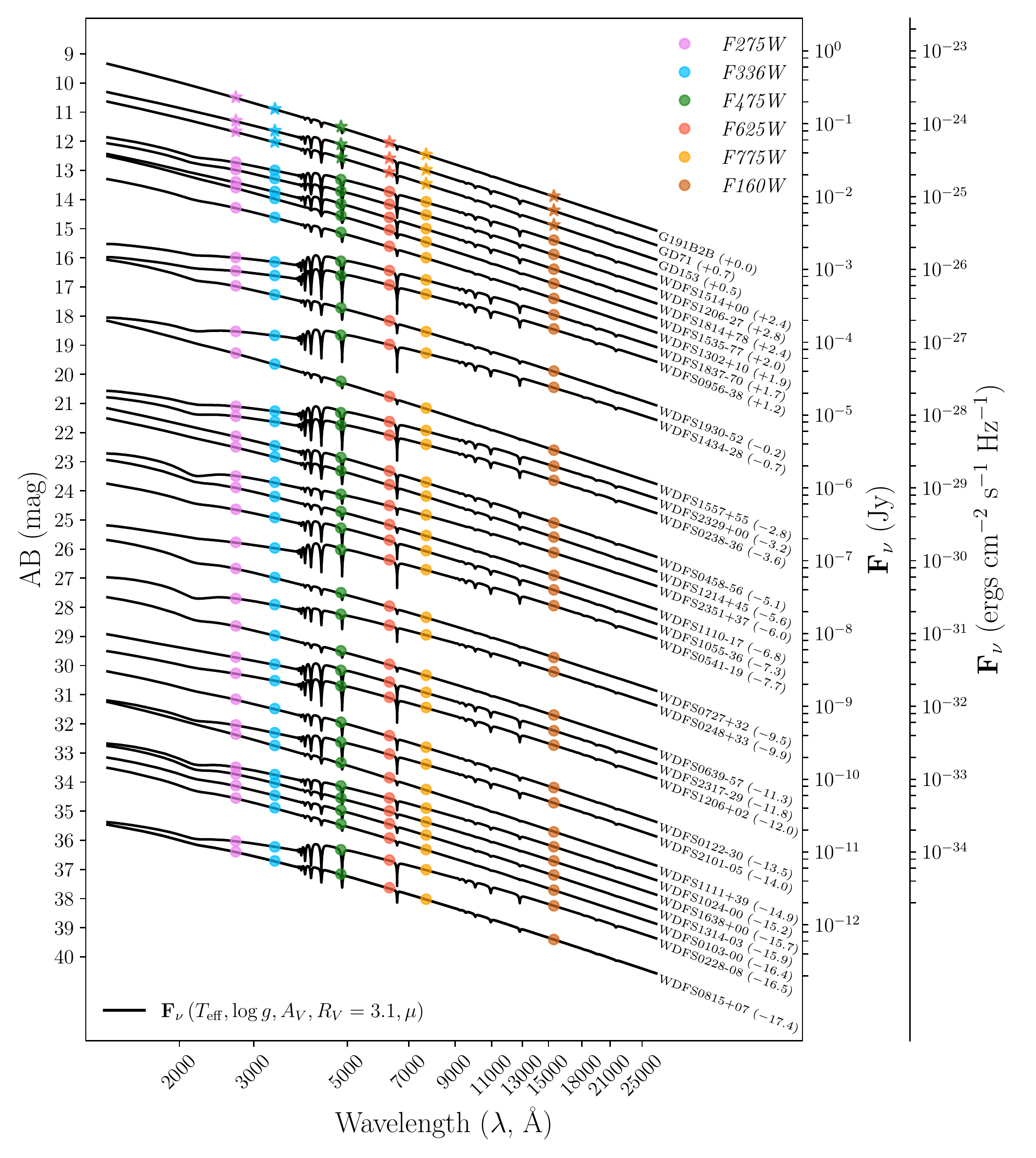}
    \caption{Calculated SEDs for all DAWDs in our network with synthetic HST magnitudes overlaid (colored points). Each spectrum is arbitrarily shifted in AB mag for clarity.}
    \label{fig:SED}
\end{figure*}
\clearpage

\begin{figure*}[hptb]
   \centering
    \includegraphics[width=\textwidth]{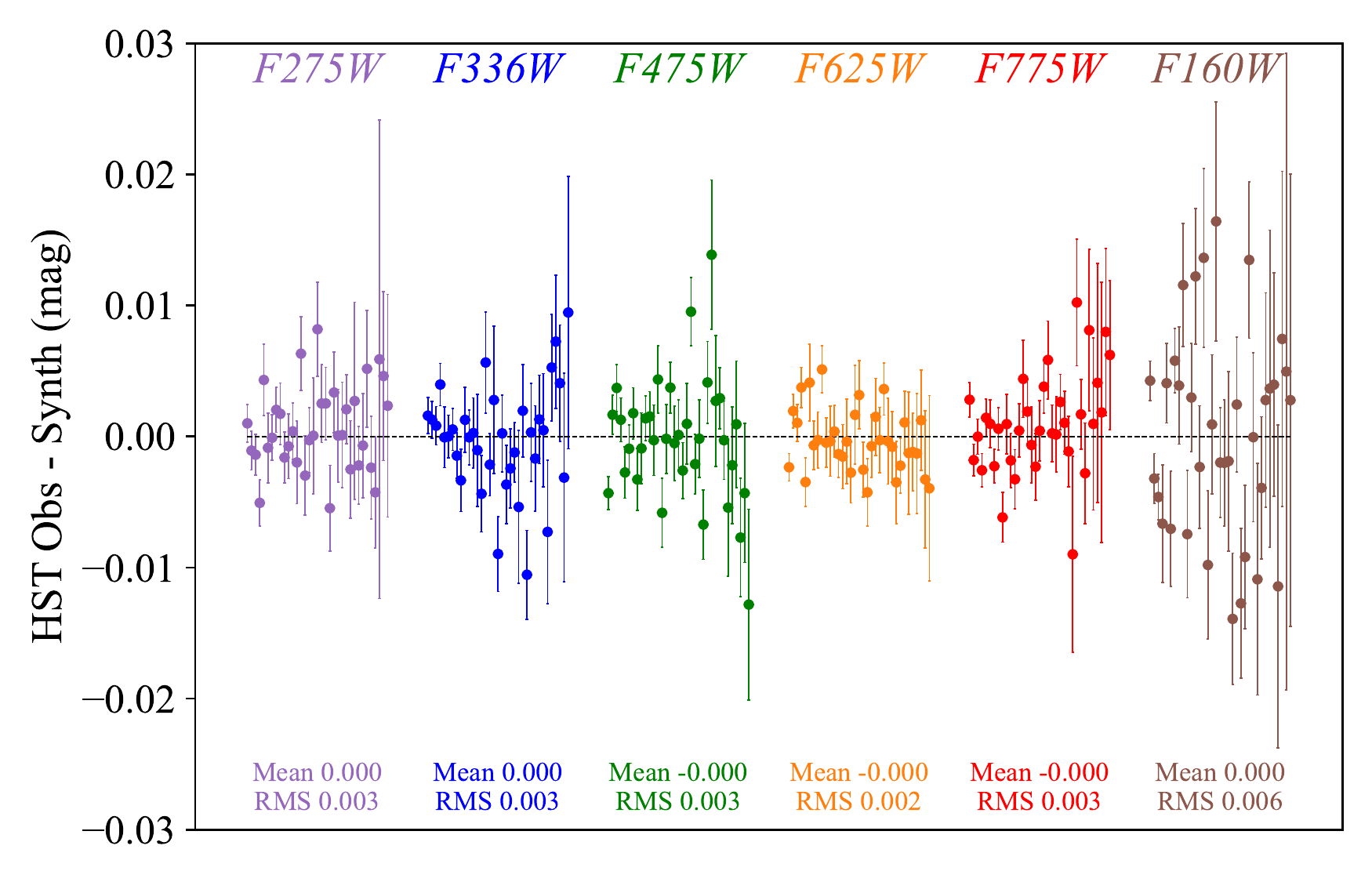}
    \caption{HST photometric residuals (in the sense of corrected observed magnitude $-$ model magnitude) for our network of DA white dwarf stars.  X axis coordinates within each band are uniformly spaced, ordered by the $F275W$ magnitude with the brightest at the left.  The mean and RMS are weighted by the photometric uncertainties.}
    \label{fig:WDmodelPhotResid}
\end{figure*}

\begin{figure*}[hptb]
   \centering
   \includegraphics[width=\textwidth]{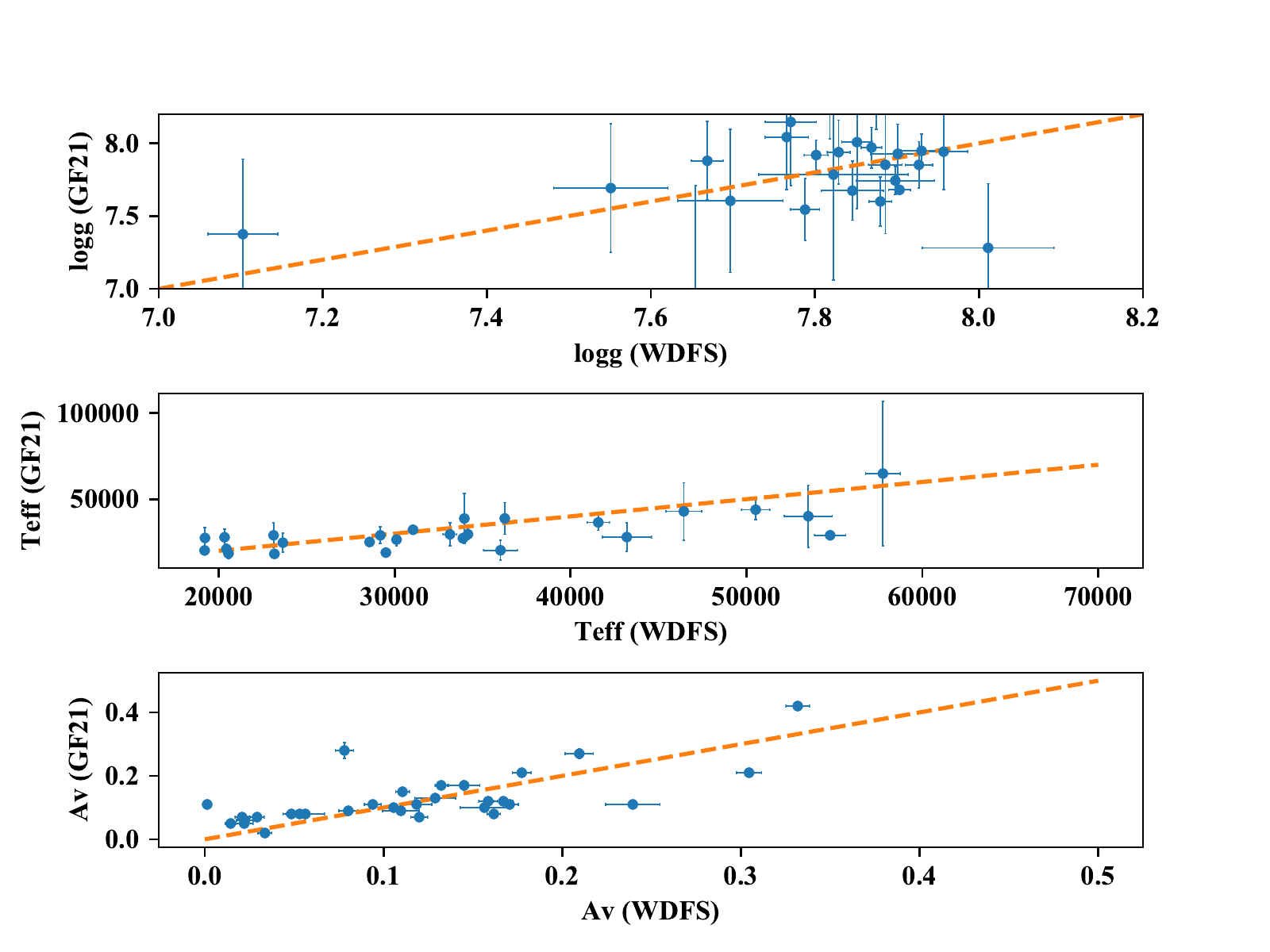}
\caption{Comparison between our values for \logg \, \teff \, and \av \, and those from \citet{Gentile-Fusillo21}, denoted in the plots as GF21. The dotted lines show the identity relation.  Note that a point in the \logg \, plot at approximate value 9.0 is off scale. }
\label{fig:GF21compare}
\end{figure*}

\begin{figure*}[hptb]
   \centering
    \includegraphics[width=\textwidth]{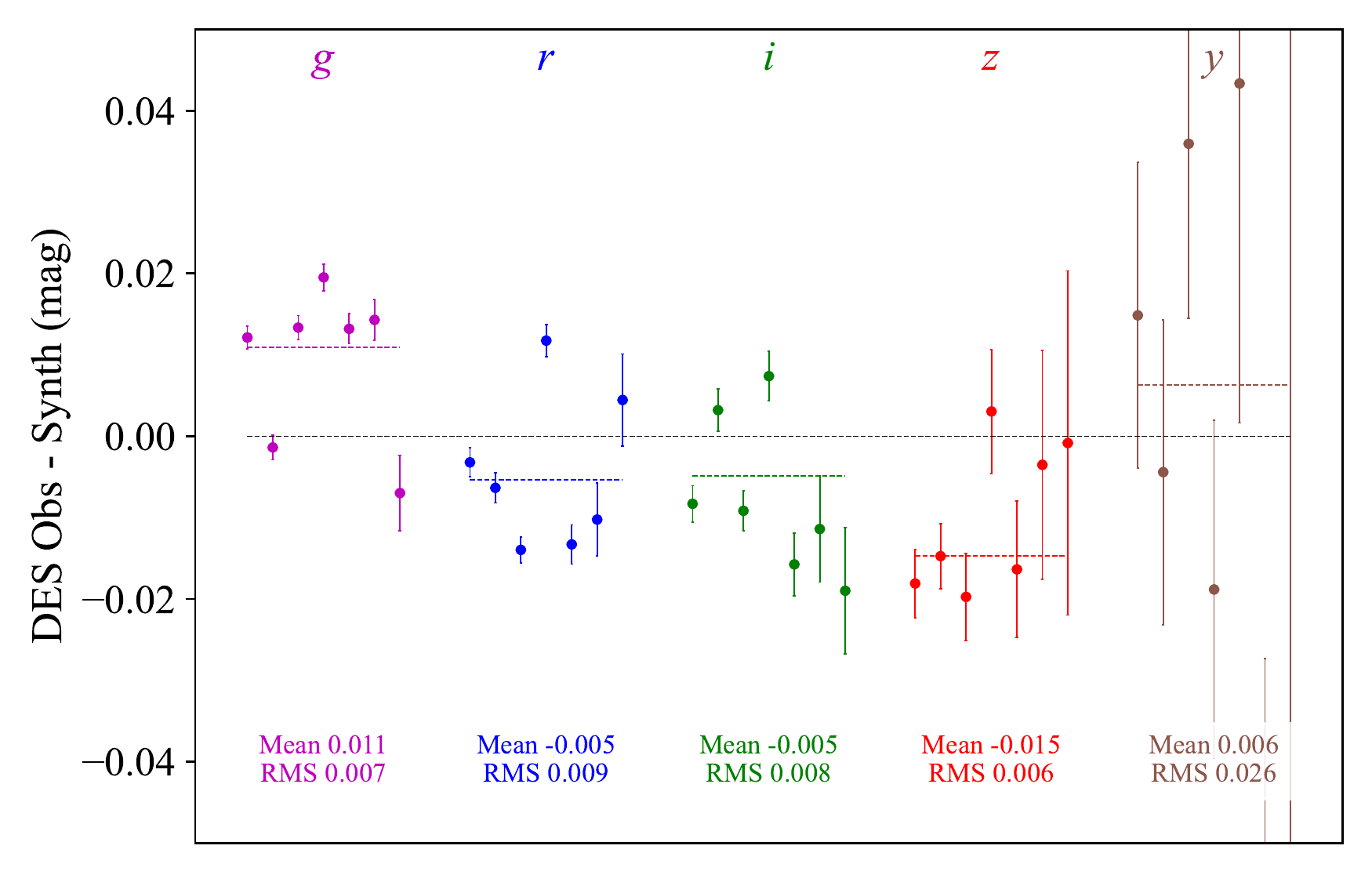}
\caption{DES observed minus synthetic magnitudes in $grizy$.  X-axis coordinates within each band are uniformly spaced, ordered by the $g$ magnitude with the brightest at the left.  The mean and RMS are weighted by the photometric uncertainties.  The black dashed line indicates zero difference.  The mean value for each filter is represented by a dashed line through those filter’s points in the same color.  Note that some points are off scale for this figure.  See Table \ref{tab:DECAM_Residuals}.}
\label{fig:DECAMResid}
\end{figure*}

\begin{figure*}[hptb]                                       
   \centering
    \includegraphics[width=\textwidth]{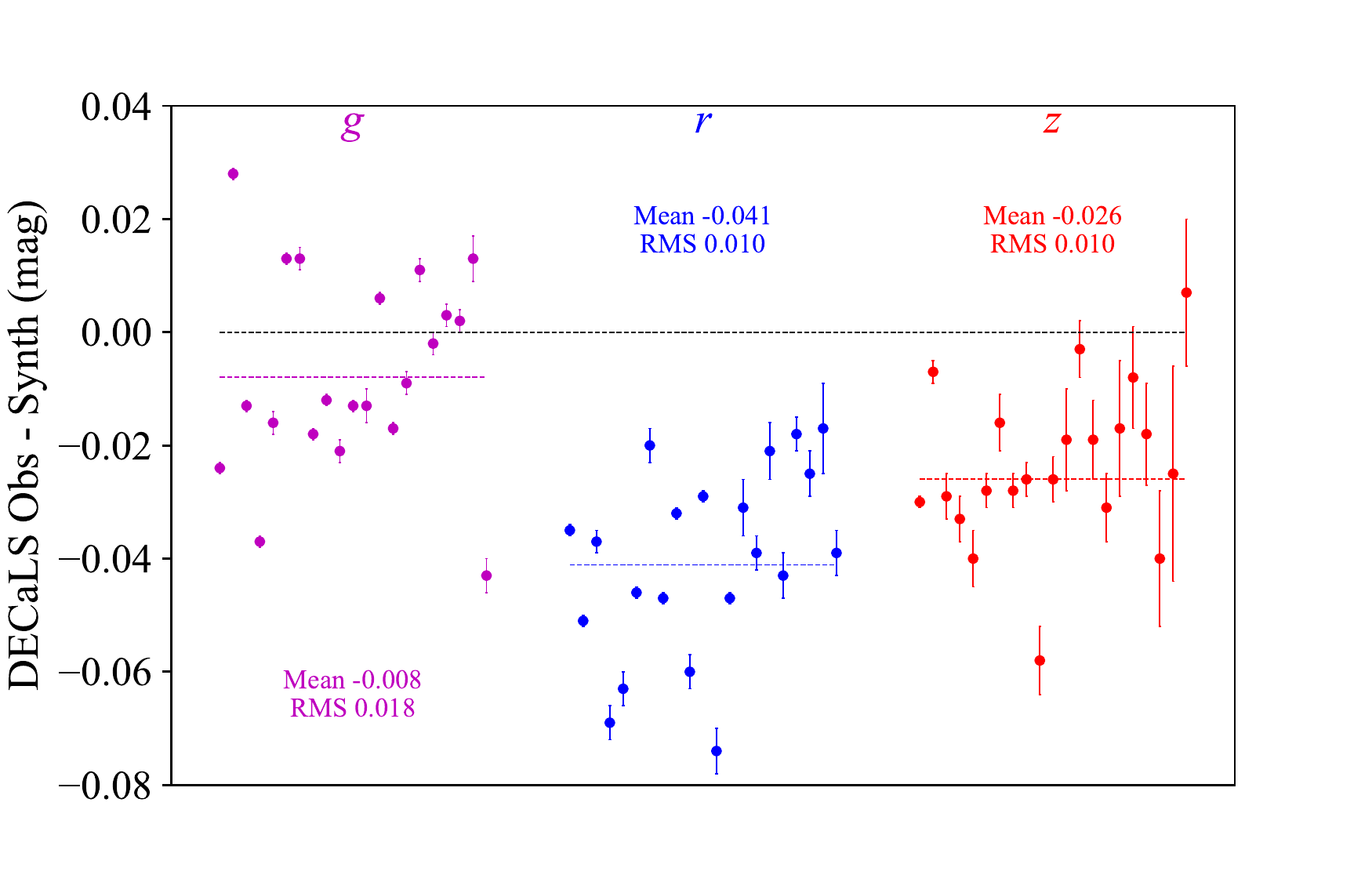}
\caption{DECaLS Survey observed minus synthetic magnitudes in $grz$.  X-axis coordinates within each band are uniformly spaced, ordered by the $g$ magnitude with the brightest at the left.  The mean and RMS are weighted by the photometric uncertainties.  The black dashed line indicates zero difference.  The mean value for each filter is represented by a dashed line through those filter’s points in the same color. See Table  \ref{tab:LS_Residuals}.}
\label{fig:LS_Resid}
\end{figure*}

\begin{figure*}[hptb]
   \centering
   \includegraphics[width=\textwidth]{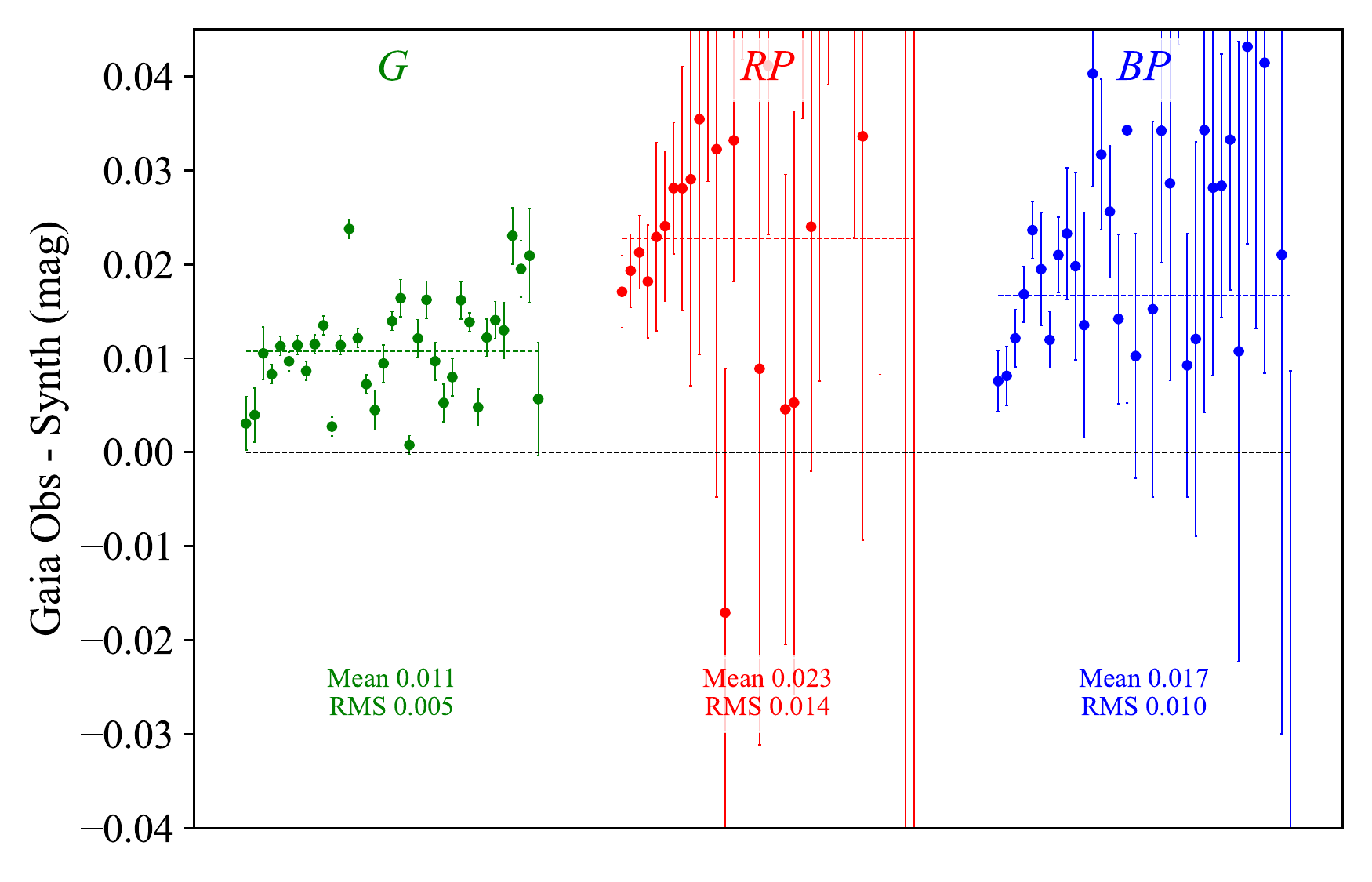}
\caption{Gaia DES observed minus synthetic magnitudes in $G$, $RP$, and $BP$.  X-axis coordinates within each band are uniformly spaced, ordered by the $G$ magnitude with the brightest at the left.  The mean and RMS are weighted by the photometric uncertainties.  The black dashed line indicates zero difference.  The mean value for each filter is represented by a dashed line through those filter’s points in the same color.  See Table  \ref{tab:Gaia_Residuals}.}
\label{fig:GaiaResid}
\end{figure*}

\begin{figure*}[hptb]
   \centering
   \includegraphics[width=\textwidth]{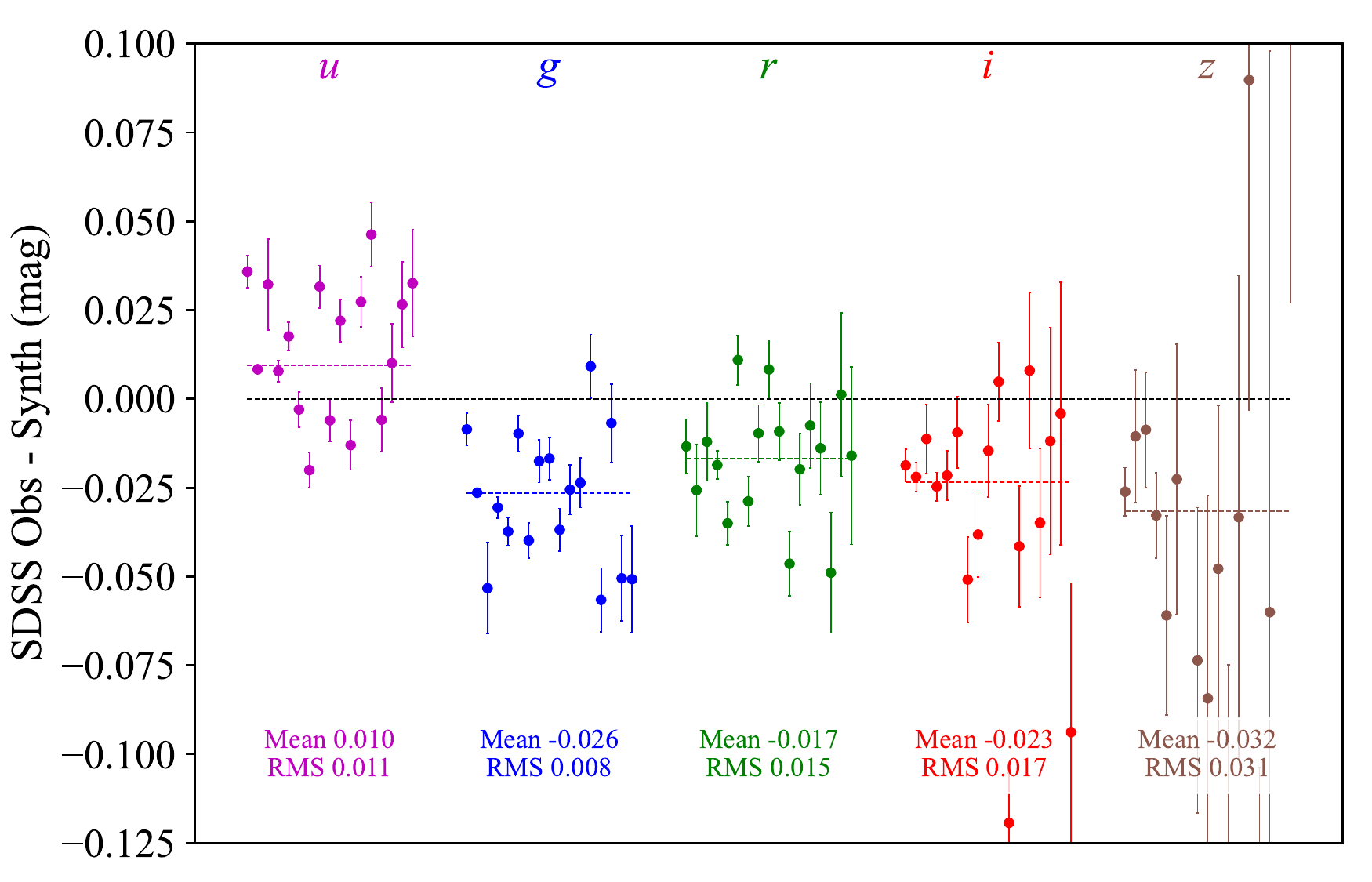}
\caption{SDSS observed minus synthetic magnitudes in $ugrizy$.  X-axis coordinates within each band are uniformly spaced, ordered by the $g$ magnitude with the brightest at the left.  The mean and RMS are weighted by the photometric uncertainties.  The black dashed line indicates zero difference.  The mean value for each filter is represented by a dashed line through those filter’s points in the same color.  Note that some points are off scale for this figure.  See Table  \ref{tab:SDSS_Residuals}.}
\label{fig:SDSSResid}
\end{figure*}

\begin{figure*}[hptb]
   \centering
   \includegraphics[width=\textwidth]{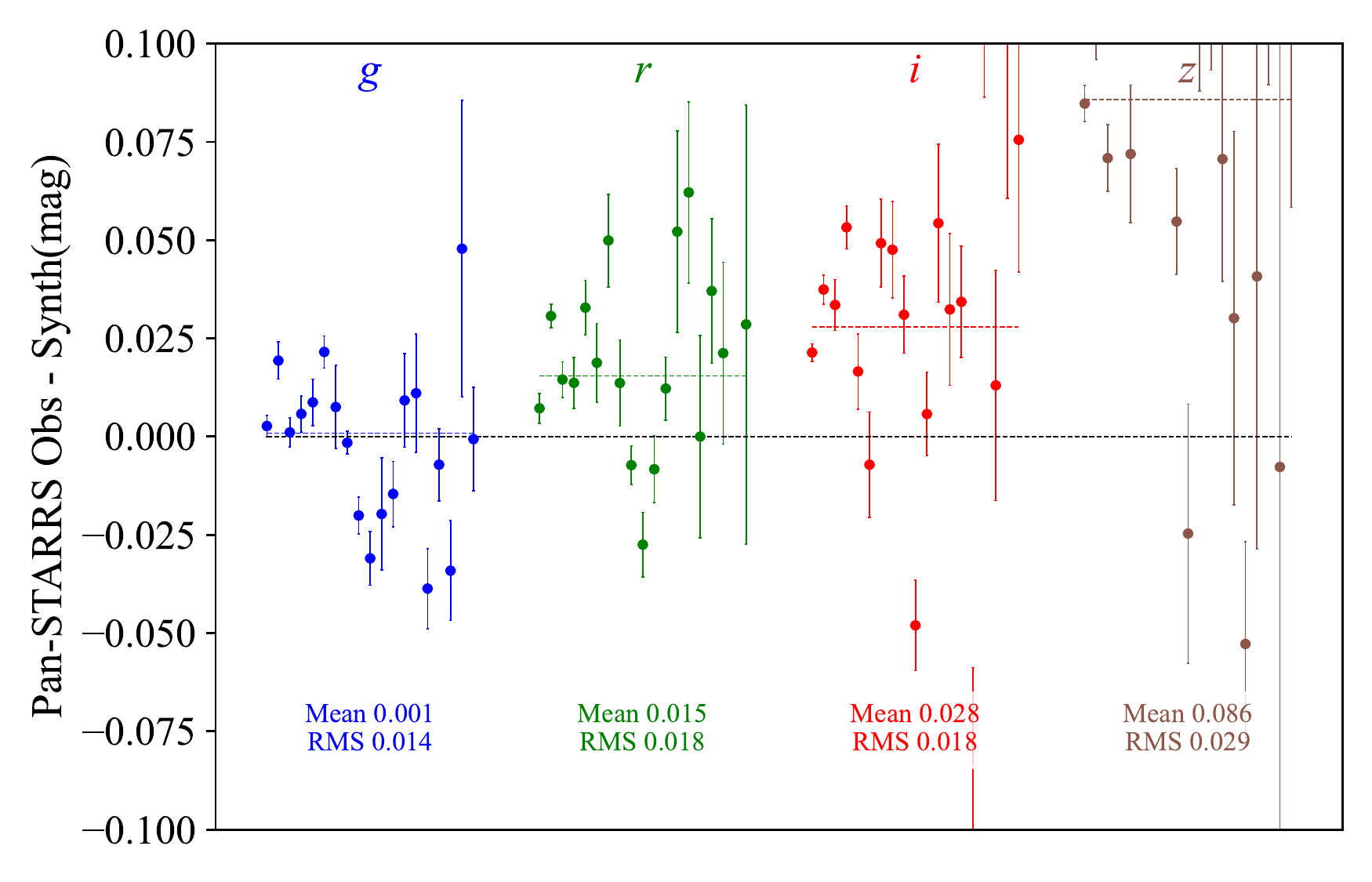}
\caption{Pan-STARRS1 observed minus synthetic magnitudes in $griz$.  X-axis coordinates within each band are uniformly spaced, ordered by the $g$ magnitude with the brightest at the left.  The mean and RMS are weighted by the photometric uncertainties.  The black dashed line indicates zero difference.  The mean value for each filter is represented by a dashed line through those filter’s points in the same color.  Note that some points are off scale for this figure. See Table  \ref{tab:PS1_Residuals}.}
\label{fig:PSResid}
\end{figure*}

\begin{figure*}[hptb]
   \centering
   \includegraphics[width=\textwidth]{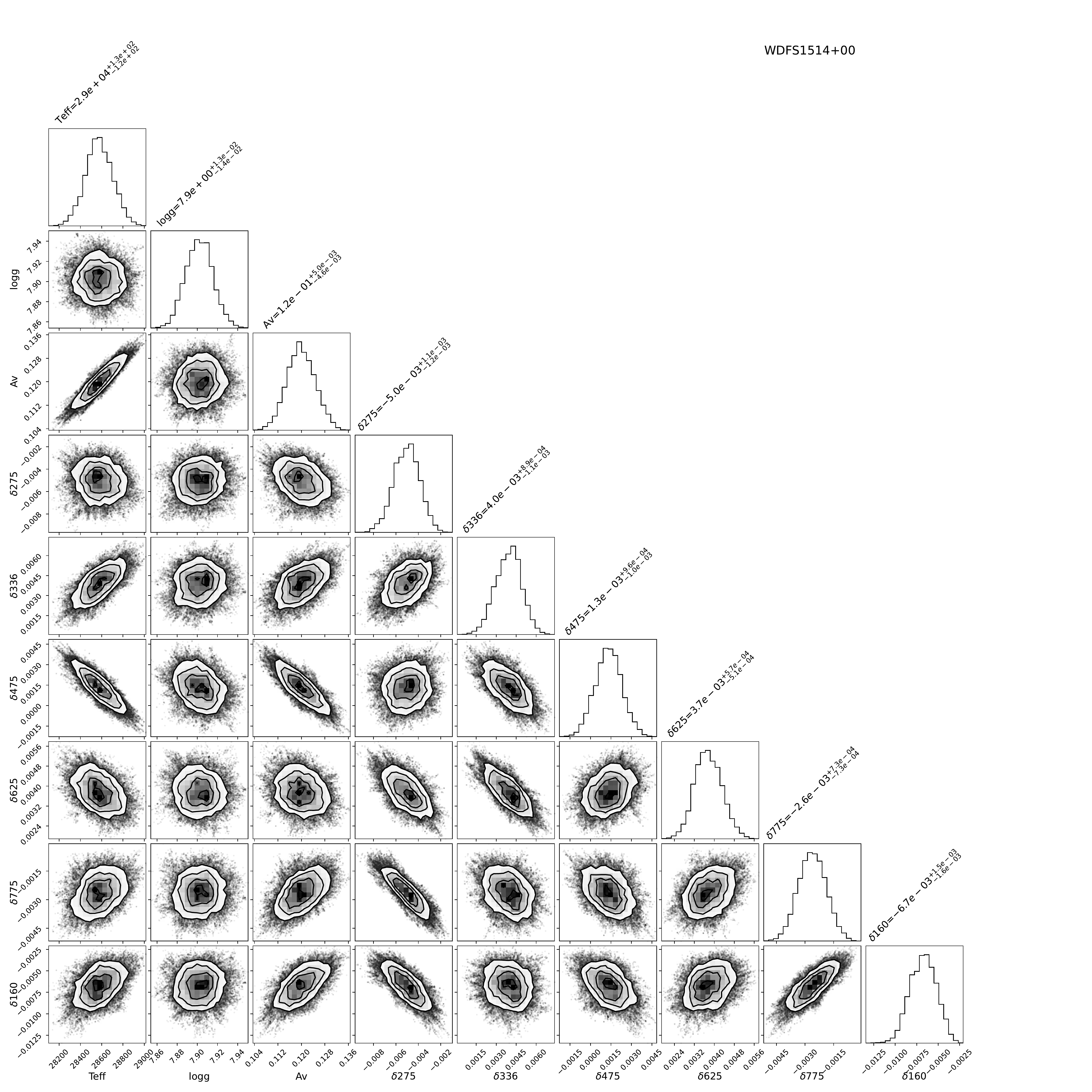}
\caption{Corner plot showing the posterior distribution of the model parameters for WDFS1514+00. $\delta275$ etc are the residuals as in Table \ref{tab:HST_Residuals}. Note the scales.}
\label{fig:corner}
\end{figure*}

\begin{figure*}[hptb]
   \centering
   \includegraphics[width=\textwidth]{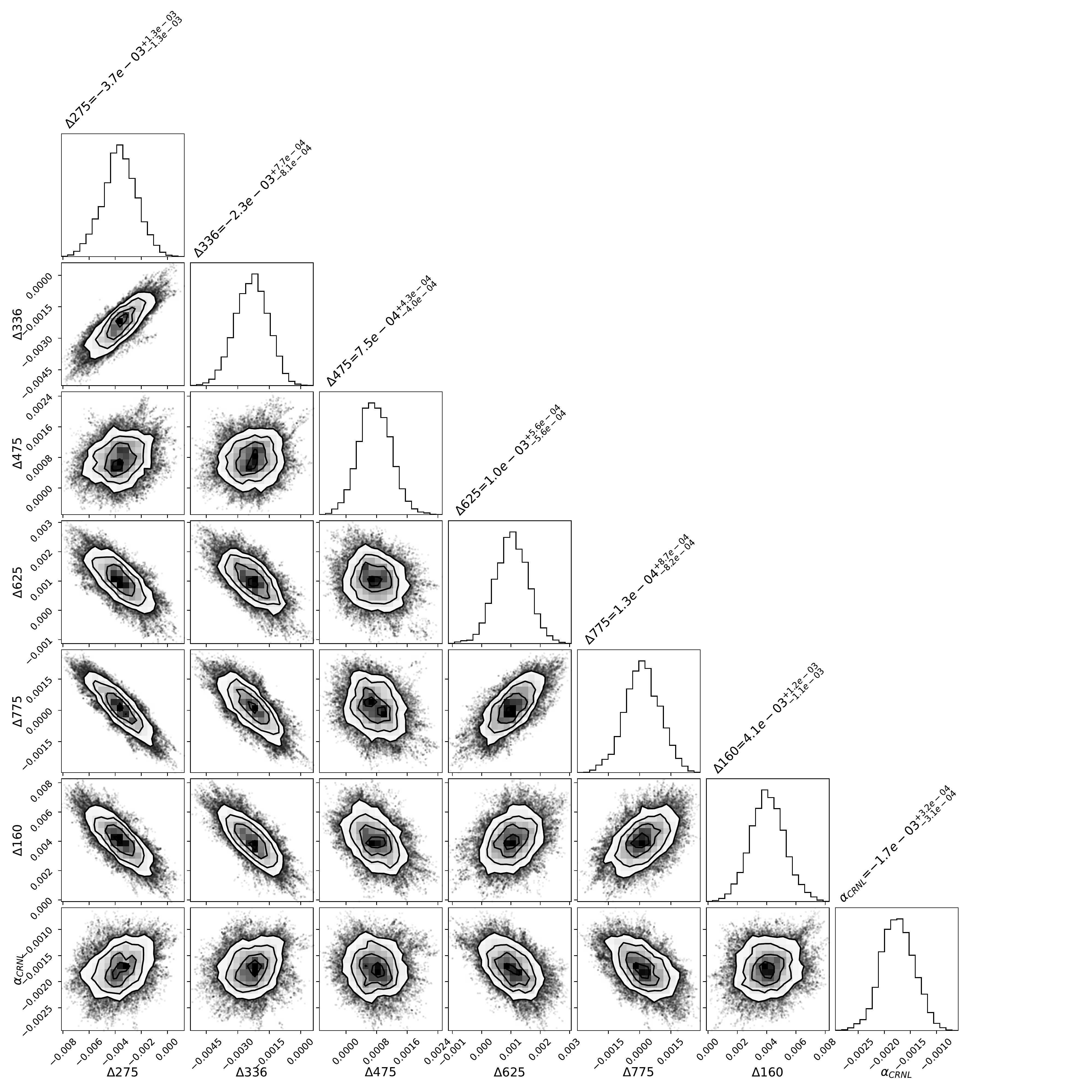}
\caption{Corner plot showing the posterior distribution of the model parameters for $\Delta_{\lambda}$ and $\alpha_{CRNL}$. Note the scales.}
\label{fig:cornerCRNL}
\end{figure*}

\begin{figure*}[h]                                                  
   \centering
   \includegraphics[width=\textwidth]{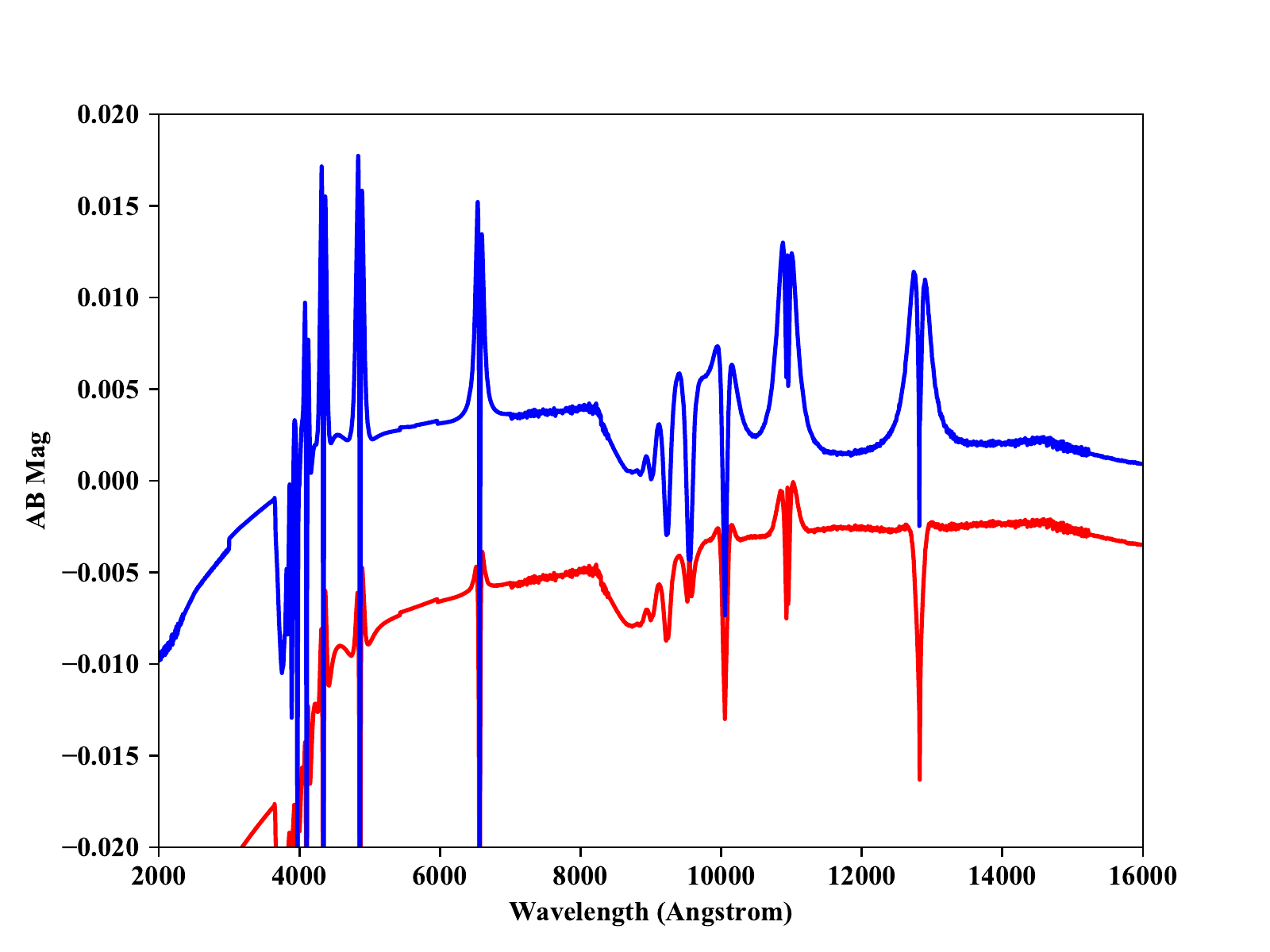}

\caption{Comparison between our model SED for GD153 and that of two CALSPEC SEDs. The blue curve shows the AB magnitude difference between the SED for GD153 from the version of CALSPEC (gd153\_mod\_010.fits) in use till 2019 and that derived in this paper. The red curve shows the same difference, but for the 2020 CALSPEC.  The 2020 CALSPEC SED uses newer atmosphere models than the 2019 SED, with an additional change in the calibration of the achromatic absolute magnitude zero-point making the CALSPEC 2020 fluxes brighter by 0.0087 mag.
}
\label{fig:CALSPECGD153}
\end{figure*}

\end{document}